\documentclass[manuscript]{aastex}
\usepackage{emulateapj5}
\usepackage{rotating}
\usepackage{graphicx}

\usepackage{graphicx,epstopdf,textcomp}

\newcommand{\bjdtdb}{\ensuremath{\rm {BJD_{TDB}}}}

\newcommand{\ms}{m s$^{-1}$}

\slugcomment{Accepted for the Astrophysical Journal}

\shorttitle{Eclipses of HAT-P-20b}
\shortauthors{Deming et al.}

\begin{document}

\title{Spitzer Secondary Eclipses of the Dense, Modestly-irradiated, Giant 
Exoplanet HAT-P-20b Using Pixel-Level Decorrelation}

%%%%%%%%%%%%%%%AUTHORS%%%%%%%%%%%%%%%
\author{Drake~Deming\altaffilmark{1,2}, Heather~Knutson\altaffilmark{3}, Joshua~Kammer\altaffilmark{3}, Benjamin~J.~Fulton\altaffilmark{4}, 
 James~Ingalls\altaffilmark{5}, Sean~Carey\altaffilmark{5}, Adam Burrows\altaffilmark{6}, Jonathan~J.~Fortney\altaffilmark{7}, 
  Kamen~Todorov\altaffilmark{8}, Eric~Agol\altaffilmark{9,2}, Nicolas~Cowan\altaffilmark{10}, Jean-Michel~Desert\altaffilmark{11}, 
  Jonathan~Fraine\altaffilmark{1}, Jonathan Langton\altaffilmark{12}, Caroline~Morley\altaffilmark{7}, and Adam~P.~Showman\altaffilmark{13}  } 

\altaffiltext{1}{Department of Astronomy, University of Maryland, College Park, MD 20742, USA; ddeming@astro.umd.edu}
\altaffiltext{2}{NASA Astrobiology Institute's Virtual Planetary Laboratory}
\altaffiltext{3}{Division of Geological and Planetary Sciences, California Institute of Technology, Pasadena, CA 91125, USA}
\altaffiltext{4}{Institute for Astronomy, University of Hawaii at Manoa, Honolulu, HI 96822, USA}
\altaffiltext{5}{Spitzer Science Center, MS\,314-6, California Institute of Technology, Pasadena, CA 91125, USA}
\altaffiltext{6}{Department of Astrophysical Sciences, Princeton University, Princeton, NJ 08544-1001, USA}
\altaffiltext{7}{Department of Astronomy and Astrophysics, University of California, Santa Cruz, CA 95064, USA}
\altaffiltext{8}{Institute for Astronomy, ETH Zurich, Zurich, Switzerland}
\altaffiltext{9}{Department of Astronomy, Box 351580, University of Washington, Seattle, WA 98195, USA}
\altaffiltext{10}{Department of Physics and Astronomy, Amherst College, Amherst, MA 01002-5000, USA}
\altaffiltext{11}{CASA, Department of Astrophysical and Planetary Sciences, University of Colorado, 389-UCB, Boulder, CO 80309, USA}
\altaffiltext{12}{Department of Physics, Principia College, Elsah, IL 62028, USA}
\altaffiltext{13}{Department of Planetary Sciences and Lunar and Planetary laboratory, University of Arizona, Tucson, AZ 85721, USA}
%%%%%%%%%%%%%%%ABSTRACT%%%%%%%%%%%%%%%
\begin{abstract} 

  HAT-P-20b is a giant metal-rich exoplanet orbiting a metal-rich
  star. We analyze two secondary eclipses of the planet in each of the
  3.6- and 4.5\,$\mu$m bands of Warm Spitzer. We have developed a
  simple, powerful, and radically different method to correct the
  intra-pixel effect for Warm Spitzer data, which we call pixel-level
  decorrelation (PLD). PLD corrects the intra-pixel effect very
  effectively, but without explicitly using - or even measuring - the
  fluctuations in the apparent position of the stellar image.  We
  illustrate and validate PLD using synthetic and real data, and
  comparing the results to previous analyses.  PLD can significantly
  reduce or eliminate red noise in Spitzer secondary eclipse
  photometry, even for eclipses that have proven to be intractable
  using other methods. Our successful PLD analysis of four HAT-P-20b
  eclipses shows a best-fit blackbody temperature of $1134\pm29$K,
  indicating inefficient longitudinal transfer of heat, but lacking
  evidence for strong molecular absorption.  We find sufficient
  evidence for variability in the 4.5\,$\mu$m band that the eclipses
  should be monitored at that wavelength by Spitzer, and this planet
  should be a high priority for JWST spectroscopy.  All four eclipses
  occur about 35 minutes after orbital phase $0.5$, indicating a
  slightly eccentric orbit.  A joint fit of the eclipse and transit
  times with extant RV data yields $e\cos{\omega}= 0.01352
  ^{+0.00054}_{-0.00057}$, and establishes the small eccentricity of
  the orbit to high statistical confidence.  HAT-P-20b is another
  excellent candidate for orbital evolution via Kozai migration or
  other three-body mechanism.

\end{abstract}

%%%%%%%%%%%%%%%INTRODUCTION%%%%%%%%%%%%%%%
\section{Introduction}
The transiting exoplanet HAT-P-20b occupies a unique niche in
parameter space, being a massive ($M=7.246\pm 0.187\,M_J$), high density planet ($\rho =
13.8\pm1.5\,g\,cm^{-3}$), orbiting a relatively small metal-rich star
($R=0.69\pm0.02R_{\odot}, [Fe/H]=+0.35\pm0.08$ \citealp{bakos}).  The
high metallicity of the star, and the radius of the massive planet being
smaller than Jupiter ($R_p=0.867\pm 0.033 R_J$, \citealp{bakos}),
suggest that the planet is metal-rich.  Moreover, HAT-P-20b is only
moderately irradiated, with a predicted equilibrium temperature of
970K (for zero albedo and uniform longitudinal distribution of heat).

Spitzer observations of exoplanets like HAT-P-20b, with equilibrium
temperatures below 1000K, are targets of a new Spitzer program
(H. Knutson, P.I.) to search for a plausible inverse relation between
planetary mass and atmospheric metallicity.  That relation should be
especially obvious at temperatures where methane forms ($\lessapprox
1000$K), because methane abundance can be supressed by CO formation
when the atmospheric metallicity is very high \citep{moses}.  We
therefore anticipate the possibility of strong molecular absorption,
due to either methane or CO, in cool giant exoplanets like HAT-P-20b.
Indeed, Spitzer eclipses of the moderately irradiated exo-Neptune
GJ\,436b are interpreted as exhibiting strong CO absorption, and a
large depletion of methane (\citealp{stevenson10, lanotte}).

Although we seek the molecular absorption spectra of giant transiting
planets, it has been argued that the emergent spectra of some hot
Jupiters - such as the strongly-irradiated WASP-12b - are consistent
with that of a blackbody \citep{crossfield}. However,
\citet{stevenson14} concludes that the spectrum of WASP-12b deviates
decisively from a blackbody.  Recently, \citet{hansen} have suggested
that the dayside spectra of {\it all} close-in giant transiting
planets are adequately described as blackbodies, based on their
inferred level of systematic error in the Spitzer analyses.
Irrespective of this debate, there are already multiple examples in
the literature where the emergent dayside spectra of hot Jupiters are
consistent with that of a blackbody.  For example, Corot-1b
\citep{deming11}, WASP-48b and HAT-P-23b \citep{orourke} resemble
blackbodies.  However, there are also planets whose emergent spectrum
is clearly not a blackbody, WASP-43b being the most recent example
\citep{kreidberg}.

Our observations of HAT-P-20b have two motivations.  First, we want to
detect molecular absorption or emission in the non-blackbody spectrum
of a moderately-irradiated metal-rich giant exoplanet.  Second, we
aspire to improve techniques for hot Jupiter photometry, by reducing
the level of systematic error in Warm Spitzer analyses. To that end,
we have developed a simple, powerful, and radically different new
methodology for producing high quality photometry from Spitzer images
at 3.6 and 4.5\,$\mu$m.  In this paper, we report a two-point
photometric `spectrum' of HAT-P-20b, based on two eclipses observed in
each Warm Spitzer band, and we discuss the implications for the
atmosphere and orbit of the planet.

This paper is organized as follows.  In Sec.~2, we introduce our new
pixel-level decorrelation (PLD) technique for analyses of Spitzer
data, and we argue for its advantages over traditional methods. Sec.~3
explains how we implement PLD in practice, and Sec.~4 validates it by
applying it to synthetic data where the underlying transit, eclipse,
and phase curve amplitudes are known.  Sec.~5 describes application of
PLD to real data for several exoplanetary systems previously analyzed
using traditional methods, and we compare our PLD results and error
levels to those previous analyses.  Having thus validated PLD as an
effective tool for Spitzer analyses, we apply it to HAT-P-20b.  Sec.~6
describes our new observations, and the initial processing of the
data. Sec.~7 derives the PLD eclipse amplitudes of HAT-P-20b, and
Secs.~8 and~9 discuss the implications of our results for the
atmosphere and orbit of HAT-P-20b, respectively. Sec.~10 summarizes
our conclusions.

%%%%%%%%%%%%%%%OBSERVATIONS%%%%%%%%%%%%%%%
\section{The Zen of Intra-pixel Decorrelation}

We here motivate and describe our new PLD technique, which differs
fundamentally from all other methods used to analyze Spitzer
data to date.  

Photometry of IRAC images at 3.6 and 4.5\,$\mu$m has long been known
\citep{charbonneau} to exhibit a systematic effect due to intra-pixel
sensitivity variations \citep{ingalls}.  When coupled with pointing
jitter, the intra-pixel sensitivity variations produce intensity
fluctuations that must be removed from photometry in order to detect
the subtle eclipses of exoplanets.  Current methods to correct Spitzer
photometry are based on defining and removing a correlation between
apparent intensity fluctuations of the host star, and its physical
position on the detector as determined by finding the centroid of the
stellar PSF.  The earliest such decorrelations (e.g.,
\citealp{charbonneau, knutson08, machalek}) modeled the intensity
fluctuations as polynomial (typically quadratic) functions of the
Y-coordinate, sometimes with a weaker (e.g., linear) dependence on the
X-coordinate.  Polynomial decorrelations are still used (e.g.,
\citealp{shporer}), but methods have evolved to include very powerful
implementations such as Bi-Linear Interpolated Sub-pixel Sensitivity
(BLISS) mapping \citep{stevenson12}, and novel variants such as a
spatial weighting-function approach \citep{ballard, cowan12}, and
modifications thereof \citep{lewis, zellem, lanotte}.  These
decorrelations have been largely, but not entirely successful.  Their
success is illustrated by very precise observations such as the
transits of GJ1214b \citep{fraine, gillon}.  However, their limitation
is indicated by residual red noise that is often found, especially in
the 3.6\,$\mu$m band where the intra-pixel effect is strongest.

All current methods to remove Spitzer's intra-pixel effect rely on
defining a relationship between photometric fluctuations and the
position of the stellar image on the detector.  However, PLD neither defines
nor requires any functional relation between intensity fluctuations
and the position of the stellar image. Although we determine the
position of the stellar image in order to measure its intensity within a
circular aperture (i.e., do photometry), we do {\it not} use the
image position {\it per se} to correct the intensity fluctuations.
We assert a zen-like irony: the best way to correct the effect of
fluctuations in image position does not involve using the position of
the image.

We point out that the position of the stellar image is a secondary
data product, derived from the intensities registered by the pixels of
the detector.  The intensities of {\it individual pixels} are the
primary data.  Conventional methods use the pixel intensities to
define an image position via a numerical process (e.g., determining
center-of-light or fitting a 2-D Gaussian). The centroid position of
the star is then related to intensity fluctuations by a second
numerical process (e.g., BLISS mapping, or polynomial fits). In
defining the position-intensity relation, the star is implicitly
treated as a point source, when in fact it has a width comparable to
the pixels it is traversing.  PLD omits these two intermediate
numerical steps, and relates the fluctuations in total intensity to
the individual pixels directly, using a simple, physically-motivated,
linear expression, as we now describe.

For most Spitzer secondary eclipse observations, the star moves by
less than one pixel over the entire time series.  Positional stability
of Spitzer images has been greatly facilitated by reductions in
pointing jitter \citep{grillmair}, and reproducibility of target
acquisition \citep{ingalls}.  A relatively few pixels typically
encapsulate most of the information concerning the total brightness and
position of the stellar image.  As the image moves (for example) in
the +Y direction, the pixel immediately above the centroid receives a
greater proportion of the total flux, and the pixel immediately below
the centroid receives less of the total flux.  The position of the
image is thus encoded in the relative intensities of the
pixels.  Hence, PLD uses positional information implicitly, but not
explicitly.  We consider a small group of pixels that contain the
stellar image, typically a $3 \times 3$-pixel square approximately
centered on the star.  Indexing the 2-dimensional grid of $N$ pixels
using a single index, let the background-subtracted intensity of pixel
$i$ at time $t$ be denoted $P_{i}^{t}$.  The measured brightness of
the star, $S^t$, in a frame of data at time $t$ can be written as:

\begin{equation}
S^t = {\mathcal{F}}( P_{1}^{t}, P_{2}^{t}, P_{3}^{t}, ... P_{N}^{t} ),
\end{equation}

where $\mathcal{F}$ is a generalized function.  Because the PSF of the
telescope is broader than an individual pixel, $S^t$ varies smoothly
with the position of the image.  In that physical situation,
$\mathcal{F}$ is continuous and differentiable, and we can apply a
Taylor series expansion to derive an expression for the fluctuations
in $S^t$ as a function of the changes in $P_{i}^{t}$.  For small
fluctuations in image position, we can approximate the Taylor
expansion using only the linear terms:

\begin{equation}
\delta S^t = \sum_{i=1}^N  \frac{ \partial  {\mathcal{F}}  } {\partial {P}_{i}^{t}} {\delta  {P}_{i}^{t}, }
\end{equation}

where the lower case $\delta$ indicates the fluctuations in $S^t$
caused only by the combination of image motion and spatial
inhomogeneities of the detector. To utilize Eq.~(2) in actual data
analysis, we first normalize the pixel intensities so that their sum
is unity at each time step, thus:

\begin{equation}
\hat{P}_{i}^{t} = \frac {P_{i}^{t}} {\sum_{i=1}^N P_{i}^{t}} 
\end{equation}

Note that the $\hat{P}$ values do not contain the eclipse of the
planet, because astrophysical variations are removed by the
normalization. Now we include purely temporal variations in detector
sensitivity, and the eclipse itself.  Both of these effects will be
mutiplicative times $\delta S^t$, and represented by factors close to
one, that can be written as $1\pm{\epsilon(t)}$ for the (small)
temporal effects, and $1-D E(t)$ for the eclipse, where D is the
eclipse depth and $E(t)$ is the eclipse shape normalized to unit
amplitude.  Multiplying Eq.(2) by these factors will produce
cross-terms such as $DE(t)c_i{\delta}\hat{P}^t_i$ (see below for
$c_i$).  The cross terms are second order and can be neglected.  For
example, $D E(t) \sim 0.001$ for the eclipses analyzed in this paper,
and $c_i \delta\hat{P}^t_i \sim 0.004$, so their product (4 parts per million)
is not significant to the data analysis.  To characterize and remove
the intra-pixel effect, while simultaneously solving for the amplitude
of the eclipse, and temporal baseline effects, we re-write Eq.(2) as:

\begin{equation}
\Delta S^t = \sum_{i=1}^N c_{i}\hat{P}_{i}^{t} + D E(t) + ft + gt^2 + h,
\end{equation}

where the upper case $\Delta$ indicates the total fluctuations from
all sources, $c_{i}$ represent the partial derivatives from the Taylor
expansion, and we added an offset constant ($h$).  We here represent
the temporal variations ($1\pm{\epsilon}(t)$) using a quadratic
function of time ($ft + gt^2$).  An exponential function of time is
also possible, and we explore that in Secs.~5.1~\&~7.1.  We find that in
practice the $\delta {P}_{i}^{t}$ values from Eq.~(2) can be replaced
in Eq.~(4) by the normalized pixel values themselves (the
$\hat{P}_{i}^{t}$).  The $\hat{P}_{i}^{t}$ terms in Eq.(4) relate the
apparent fluctuations in stellar intensity to the manner in which that
intensity is distributed among the detector pixels.

We obtain $S^t$ (and thus $\Delta S^t$) from a circular numerical
aperture centered on the star.  But $S^t$ could also be derived from
the sum of pixels in Eq.(3), and we discuss this option in
Sec.~5.4. Note also, that Eq.~(3) guarantees that the
$\hat{P}_{i}^{t}$ values are not trivially related to the $\Delta
S^t$.  Moreover, there is nothing that limits PLD to using a linear
expansion in Eqs.\,(2)~\&~(4).  Non-linear terms from the Taylor
expansion (e.g., quadratic in one or more $\hat{P}_{i}^{t}$) could be
included if the physical situation warrants, i.e. if image motion is
large.  Figure~1 illustrates the principle of PLD by graphically
showing the terms that add to Eq.~4.

PLD has several major advantages over the usual method of deriving an
image position and expressing $\Delta S^t$ as some function of the
image coordinates.  The advantages of PLD are:

\begin{itemize}

\item Flat-fielding inaccuracies \citep{carey} are automatically and
  efficiently corrected by the $c_{i}$ coefficients.  When the image
  moves and a greater proportion of the stellar photons fall on a
  given (mis-calibrated) pixel, the integrated intensity could
  fluctuate in a manner poorly represented by functions adopted in
  conventional intra-pixel decorrelations. For example, if a single
  pixel has a very discordant response, the spatial effect could be
  sufficiently localized as to require a high order polynomial to
  model it, and therefore require multiple coefficients for an effect
  caused by a single pixel.  However, the PLD coefficients $c_{i}$
  each represent individual pixels one-to-one, so an efficient
  flat-fielding correction is a natural by-product of the intra-pixel
  removal.

\item PLD has a sound analytic basis: although the Taylor expansion
  (1) is only approximate in practice, it is rigorous for
  infinitesimal displacements of the image.  Moreover, the PLD
  coefficients usually reflect the obvious physical importance of any
  given pixel.  For example, small coefficients naturally occur for
  pixels that contribute little to the total flux.  MCMC posterior
  distributions can be used to eliminate unnecessary or redundant
  pixels. Pixels whose distributions of $c_i$ are consistent with zero
  are not affecting the solution, and can be dropped.  Note that
  correlations between the $c_{i}$ coefficients are physically
  expected, and are not a flaw in the procedure.  As the image moves,
  the amount of stellar flux falling in a steep sensitivity gradient
  of one pixel can be accompanied by an opposite effect for a
  neighboring pixel.  However, we find no correlations between the
  $c_i$ and the {\it eclipse depth}.

\item PLD is a very effective technique, capable of removing red noise
  that frustrates other methods. Red noise in Spitzer photometry is
  not noise {\it per se}, it is the response of the detector pixels to
  time-varying illumination. PLD is successful because it includes all
  pixels having a significant contribution to the flux, and it allows
  the pixels themselves to define the red `noise' fluctuations.  Also,
  our solutions of Eq.~(4) do not merely find the best solution on the
  time scale of a single exposure.  Rather, we explicitly consider
  longer time scales when finding the best solution, as explained in
  Sec.~3.3.

\item PLD is computationally fast; it is hard to envision a faster
  method when using MCMC.  The $\hat{P}_{i}^{t}$ are computed prior to
  initializing the MCMC, and they are used with simple linear
  coefficients.  There is no weighting function to calculate, and no
  spline interpolating (as in BLISS mapping). Calculation of the
  eclipse model is the most computationally-intensive portion of PLD,
  but that calculation is also used by all other methods.

\end{itemize}

\section{PLD Fitting and Data Binning}

Binning the data in time to various degrees is an integral part of our
PLD fitting method, for both mathematical and physical reasons.  We
bin both the aperture photometry and the $\hat{P}_{i}^{t}$ values, immediately
after calculating the photometry.

\subsection{Mathematical Motivation for Binning}

There is a purely mathematical reason for binning the data.  The
coefficients ($c_i$ in Eq.\,4) of the best fit are a function of the
bin size.  A similar statement is true for {\it all} methods that
solve for the intrapixel effect in Spitzer data, because it's a
general property of least-squares solutions, not of the method to find
the solution, and not even specific to Spitzer data.

The general problem of parametric estimation in the presence of noise
has been extensively treated in the statistical literature (e.g.,
\citealp{deming43, fuller}).  In the simplest case, independent of
Spitzer, a dependent variable (Y) varies as a linear function of an
independent variable (X).  When the measurement errors are confined to
Y, the solution having the minimum reduced $\chi^2$ does not depend on
the bin size, if the binning is done with proper weighting by the
inverse of the variance.  However, independence of the solution on bin
size does not hold in general.  It specifically does not hold when X
also contains measurement errors.  In the Spitzer case, the X
variables are either the position of the image for conventional
decorrelations, or the $\hat{P}_{i}^{t}$ for PLD.  Since those both
contain measurement errors, the best fits to Spitzer data are
intrinsically a function of bin size.  That is true even if the errors
are purely Gaussian white noise.

Binning both the photometry and the pixel values as a function of
time, our PLD regression will find a linear solution to Eq.(4) that
minimizes the $\chi^2$ for the binned data, but {\it not} for the
unbinned data.  We find that solutions based on binned data often
exhibit less noise on the time scale of the eclipse, but always have
slightly greater point-to-point scatter when those coefficients are
applied to unbinned data, versus a solution of Eq.(4) obtained on the
unbinned data directly.  Essentially, we accept greater scatter on
short time scales, as a trade-off for minimizing the noise on longer
time scales, as we explain in detail below.  Nevertheless, Eq.(4) is
sufficiently effective that our solutions often exhibit less scatter
than traditional methods on {\it all} time scales.

\subsection{Physical Motivation for Binning}

The physical reason for binning the data is related to the properties
of the Spitzer telescope.  It exhibits pointing jitter on a wide range
of time scales.  Besides the well known 40-minute oscillation due to
the battery heater, there are also short-term fluctuations from a few
to tens of seconds.  For example, the frames within a given sub-array
data cube at a 2-second cadence exhibit pointing variations that are
obvious in our photometry of both real and synthetic data.  Binning
averages out the effect of short term spatial fluctuations, and
permits the solution to focus on removing the longer-term variations
on the time scale of the planet's eclipse.  Binning is also helpful
because the pixels at the edge of the stellar PSF have relatively low
flux levels, and binning helps to improve the precision of the
$\hat{P}_{i}^{t}$ that form the basis vectors of the PLD
decorrelation.

Some consequences of binning should be mentioned.  Binning can in
principle distort the eclipse curve, and negatively affect the
solution \citep{kipping}.  However, the bin widths we use are not
sufficiently long in temporal span to produce distortion of the
eclipse curve.  We check our solutions and vary the binning to be
sure that the derived eclipse depth does not vary systematically with
bin size.  A positive effect of binning is that it helps to reduce
red noise because the binned data are more representative of lower
temporal frequencies than are the unbinned data.  We avoid binning the
data to the degree that would cause the number of data points to be
comparable to the number of coefficients that comprise the solution
(in other words, we maintain a high degree of freedom).

\subsection{A Broad-Bandwidth Solution}

We here describe specifically how we select the best PLD fit to a
given Spitzer eclipse, and determine the errors.  We perform aperture
photometry using both constant and variable radius apertures. Prior to
the binning, we solve Eq.(4) using matrix inversion repeated over a
trial grid of different central phases, to select the best-fit eclipse
phase.  (The matrix inversion finds the minimum $\chi^2$, so any other
procedure to minimize $\chi^2$ would be equivalent.) Fixing the
eclipse phase to that initial best-fit value, we vary the bin size and
again solve Eq.(4) for all combinations of bin sizes and photometry
data sets.

We use binning in two different ways.  First, there is the binning of
the photometry and the $\hat{P}_{i}^{t}$ values as described
above. For each eclipse, we consider all combinations of bin size,
photometric aperture type and size, and centroiding method.  We
explore bin sizes of 1 exposure per bin, and 2 to 258 exposures per
bin, in increments of 4 (1, 2, 6, 10, etc.).  We use 11 apertures,
and a variable radius aperture with 11 different increments added (as in
\citealp{beatty}), and two centroiding methods (2-D Gaussian fitting
and center-of-light).  At each combination, we apply the $c_i$
coefficients to the unbinned data, and calculate residuals (data minus
fit). We then explore the noise properties of those residuals using a
second binning process.  We denote the standard deviation of the
unbinned residuals as $\sigma (1)$.  We bin those residuals over 2, 4,
8, etc. points, increasing the `residual-bin' size by a factor of two,
stopping when the number of points after residual-binning is $\leq
16$.  For each residual-bin size $N$, we calculate the standard
deviation of those binned residuals $\sigma (N)$ and the $\chi^2$ of
the $\sigma (N)$ compared to a line of slope $-0.5$ that is forced to
pass through $\sigma (1)$.  The fit (bin size, aperture, centroiding
method) that minimizes the $\chi^2$ of this line is our adopted PLD
regression solution.

Our fitting criterion is a generalization of previous Spitzer
decorrelation work.  To our knowledge, all previous Spitzer
decorrelations find a best fit to unbinned data, and accept the
consequences for the residuals on longer time scales.  Our PLD fitting
exploits the mathematical and physical reality that the best fit is a function of
the time scale, i.e. the degree to which the data are binned.  By
adopting a grid of residual-bin sizes, we are considering a range of
time scales equally spaced in the logarithm of time.  Minimizing the
$\chi^2$ of our $\sigma (N)$ compared to a line of slope $-0.5$
chooses the fit that minimizes the noise over that range of time
scales, i.e., {\it we adopt a fit with broad bandwidth
  characteristics}.  That, together with the intrinsic effectiveness
of Eq.(4), allows us to greatly reduce red noise in our solutions.

After finding the best solution as described above, we use that
regression solution to initialize a Markov Chain Monte Carlo (MCMC)
procedure \citep{ford} that explores parameter space, operating on the
binned data at the degree of binning chosen by the regression. The MCMC
varies all of the eclipse parameters, including the central phase. Our
MCMC formulation uses the Metropolis-Hastings algorithm with Gibbs
sampling, and our code automatically adjusts the step size for each
parameter to converge to an acceptance rate of 0.45.  Our chains
converge and mix very quickly, because the regression solution finds
the best-fit values of the $c_{i}$ at the outset.  We confirm good
convergence and mixing (for all of the eclipses analyzed in this paper)
by comparing three independent chains, each of $10^6$ steps.  The MCMC
is sometimes able to find a slightly better solution than the
regression, but the difference is never physically significant.
Instead, the primary purpose of the MCMC is to determine the errors
and to test for correlations and degeneracies.

For each planet (real or synthetic) where we have applied PLD, we list
the properties of the best-fit solution, including the bin size used, in
Table~1.

\section{Tests of PLD Using Synthetic Data}

We have tested PLD using both synthetic and real data.  This Section
describes the tests using synthetic data.  Sec. 4.1 briefly summarizes
how the synthetic data are produced, and Secs. 4.2 and 4.3 test PLD on
two variants of the synthetic data.

\subsection{Synthetic IRAC data}

We generated and analyzed synthetic BCD files for 3.6 and 4.5\,$\mu$m,
based on a new capability developed at the Spitzer Science Center
(by. J. Ingalls \& S. Carey).  Some of these synthetic data for
WASP-52 were initially produced for the IRAC Data Challenge
Workshop\footnote{http:conference.ipac.caltech.edu/iracaas224/data-challenge}
held in association with the 224th meeting of the American
Astronomical Society.  Details of the synthetic data generation will
be published by J. Ingalls and S. Carey, but we here summarize the
essential features.

The synthetic data utilize the current best realizations of
Spitzer/IRAC's pixel sensitivity map and the telescope pointing
fluctuations. The intra-pixel effect is explicitly modeled, but
pixel-to-pixel variations in responsitivity due to flat-fielding
errors are not currently included. The interaction between the
telescope's point spread function (PSF) and the modeled intra-pixel
detector sensitivity structure is calculated for each Fowler sample of
each simulated frame.  The telescope pointing is simulated at 1 msec
time resolution, and includes fluctuations due to cycling of a heater
used to stabilize a battery in the pointing system, and a settling
drift that occurs for about 30 minutes at the start of each AOR.

The simulated observation of WASP-52b used approximately 1.3 planetary
orbits (53 hours total), divided into 12-hour AORs.  The telescope
PCRS re-acquistion was simulated for each AOR.  The total data
comprise 95,104 exposures of 2 seconds each, divided into 1486 data
cubes of dimension $32 \times 32 \times 64$.  Detector read noise and
stellar photon noise was added to each frame, but we also produced and
analyzed a noiseless version, described immediately below.  These data
also contain two spike-like fluctuations in the noise-pixel parameter,
caused by high reequency (10 Hz) pointing oscillations that smeared
out the telescope PSF when integrated over 2 seconds. These were
included to challenge the participants in the IRAC Workshop.

\subsection{Testing PLD Using Noiseless Data at 3.6 Microns}

A major advantage of synthetic data is that we can turn off the noise,
and examine the nature of the decorrelation process with maximum
clarity.  We generated synthetic noiseless data for WASP-52 at
3.6\,$\mu$m, where the intra-pixel effect is strongest.  The planet
was also turned off in this version of the data, so that we can
isolate effects of the detector.  We performed aperture photometry (on
these data as well as all of the real data in this paper) using both
apertures with constant radii from 1.6 to 3.5 pixels in increments of
0.2 pixels, and variable-radius apertures based on the noise-pixel
formulation described by \citet{lewis}. The variable-radius apertures
include a constant added to the noise pixel radius (defined by
\citealp{beatty}, $\sqrt{\beta}$, their Eq.~1), which varied from zero
to two pixels.  We located the centroid of the stellar image using
both an azimuthially symmetric 2-D Gaussian fit, as well as an
intensity-weighted center-of-light calculation in X and Y. We
decorrelated the intra-pixel effect in these data using both PLD and
polynomial fits to the X and Y positions of the image centroid.

We explored many possible combinations of constant-radius
vs. variable-radius apertures, and centroiding (Gaussian fitting
vs. center-of-light), in order to draw robust conclusions.
Centroiding affects a conventional decorrelation in two ways. First,
it determines where the photometry aperture is placed, thus impacting
the photometry. Second, it determines the X- and Y-positions of the
image that are used in the decorrelation.  For the conventional
polynomial decorrelation, we found the best results using a
variable-radius aperture and center-of-light centroiding.  The PLD
solutions do not use the image positions directly, and we found for
this case that the PLD results were relatively insensitive to choices
of centroiding and photometric apertures.  (The real data we analyze
exhibit greater sensitivity to those choices, as described in Sec.~5.)

Some results from this test are illustrated in Figure~2.  The top
panel shows the photometry for the noiseless data prior to
decorrelation.  The second panel overlays (in red) the fitted function
from the PLD regression onto the photometry.  The two lowest panels
show the residuals from the best fit for both the PLD and polynomial
cases.  Since there is no noise and no planet, all of the structure in
the photometry is due to intra-pixel detector sensitivity variations.
Neither technique removes all of the structure in the photometry, as
evidenced by non-zero residuals in the two lowest panels of Figure~2.
Certainly the detector sensitivity structure is not precisely
quadratic, so the polynomial decorrelation is substantially imperfect.
As for PLD, Eqs.\,(2)~\&~(4) are accurate only in the limit of small
changes in spatial position, and these test data exhibit relatively
large fluctuations in position (up to 0.72 pixels in Y and 0.36 pixels
in X).  Nevertheless, the standard deviation of the PLD residuals (473
parts-per-million, ppm) is sufficiently small that it would not
significantly limit most exoplanet observations, if combined in
quadrature with photon noise.  For reference, the photon noise of
WASP-8, HAT-P-20, and WASP-14, in a 2-second frame time is 2150, 2500,
and 2612 ppm, respectively.  The standard deviation of the
residuals from the polynomial fit in Figure~2 is 946 ppm, twice the
PLD value.  Using different centroiding and apertures, the polynomial
decorrelation performs even more poorly compared to the PLD result,
that is insensitive to the methodology of the photometry.  Neither
technique deals well with the noise-pixel spike due to PSF variation,
but such spikes are rare in real data.

We used a quadratic polynomial for Figure~2 because that order is
commonly used in real Spitzer decorrelations (e.g., \citealp{deming11,
  todorov12, todorov13}).  However, that's arguably an unfair
comparison because the quadratic decorrelation has only four
position-dependent parameters, vs. nine for PLD.  Therefore we also
performed solutions using only the five brightest pixels in PLD, and
comparing to polynomial decorrelations that are third and fourth order
in both $X$ and $Y$.  For the third and fourth order polynomials (6
and 8 parameters respectively), the residual error level is 601 and
553 ppm, respectively, whereas the 5-pixel PLD residual error is 511
ppm.  Hence PLD is a more efficient decorrelation method than
polynomials.  Nevertheless, we note that polynomial decorrelations
continue to be useful, for example in the recent re-analysis of
GJ\,436b \citep{lanotte}. Moreover, in Sec.~7.2 of this paper we
describe a sanity check of our HAT-P-20 results using a polynomial
decorrelation.

The Figure~2 data exhibit much larger image motion than occurs in
many, but not all, Spitzer eclipse observations.  We examined how the
amplitude of residuals in Figure~2 depends on the magnitude of the
image motion.  For image motion less than 0.03 pixels, the standard
deviation of the residuals is 163 ppm, increasing smoothly to 461 ppm
at 0.2 pixels of image motion.  Beyond 0.2 pixels of image motion, the
residual envelope varies less smoothly, but reaches 579 ppm at 0.3
pixels (not illustrated).  Less than $\sim 0.2$ pixels of image motion
is the region where our current version of PLD achieves optimum
performance, but it still exceeds the performance of polynomial
decorrelations even for image motion as large as 0.7 pixels.

We conclude that PLD produces a good fit to the intra-pixel detector
structure, at least twice as good as a quadratic polynomial
decorrelation, which is still commonly used in Spitzer analyses.
However, in actual practice that factor of two will be significantly
diluted by photon noise.  On the other hand, PLD is minimally
sensitive to the choice of centroiding and construction of the
numerical aperture used in the photometry.  (In Sec.~5.4 we show a PLD
result that does not even require measuring the position of the star.)

\subsection{Testing PLD with Synthetic Data for WASP-52b at 4.5 Microns}

We also analyzed synthetic data containing both the planet (WASP-52b)
as well as detector read noise and stellar photon noise.  These data
at 4.5 microns were analyzed by the community in the IRAC
Workshop mentioned above, except that the noise model is now revised
to properly account for the photon noise of previous Fowler samples.
They comprise the same number of 2-second exposures as the noiseless
data described above, covering a transit of the planet and two
secondary eclipses.  The planet in these data also exhibits a
sinusoidal phase curve effect.  We know in advance that the phase
curve has a minimum at the center of transit and a maximum at the
center of secondary eclipse.  We also know that the synthetic transit
occurs at phase 0.0 and uses no limb darkening.  The eclipse is
specified to occur exactly at phase 0.5, but the three amplitudes
(transit and eclipse depth, and phase curve amplitude) are unknown.

We performed photometry on these data using a similar procedure as for
the noiseless data, using both constant-radius and variable-radius
apertures.  In this case, we decorrelate the photometry using only
PLD, and we compare the retrieved amplitudes to their input values in
order to confirm that PLD is a valid method for decorrelating Spitzer
data.  Our fitting procedure is explained in Sec.~3.3.  Table~1 lists
the fitting parameters for these synthetic data, and the retrieved
amplitudes in comparison to the known values.

Figure~3 shows the photometry for WASP-52b prior to decorrelation (top
panel), as well as the decorrelated binned data with best-fit orbital
phase/transit/eclipse curve, and the residuals from the best binned
fit.  The retrieved amplitudes for the eclipse and transit parameters
(Table~1) are in excellent agreement with the input values.  Our
retrieved eclipse depth differs from the input value by $0.2\sigma$,
and the retrieved transit depth differs by $1.4\sigma$, both
consistent with random noise.  However, our retrieved phase curve
amplitude differs from the input value by $3\sigma$. Our posterior
distributions are very close to Gaussians, so that difference is very
unlikely to be due to random error. The total range of image motion in
these synthetic data exceeds 0.7 pixels, whereas Eq.(2) is only
precise in the limit of small image motion.  The real data we analyze
all have less than a third as much image motion (see Table~2), and (as
we show below) PLD produces robust results for the real eclipses. We
conclude that PLD is a valid method for analyzing Spitzer exoplanet
eclipse and transit data, but that it may require modification
(e.g. adding higher order terms to Eqs.\,2~\&~4) in order to analyze
phase curve data.

\section{Testing PLD with Real Data}

We have tested PLD with {\it five} sets of real data: GJ\,436
\citep{ballard}, CoRoT-2b \citep{deming11}, WASP-14b \citep{blecic},
WASP-8b \citep{cubillos}, and WASP-12b \citep{cowan12, stevenson14}.
These data were chosen to represent a wide range of analysis
situations. Although the purpose of these tests is primarily to
validate PLD, we also obtain new astrophysical information,
specifically a revised 3.6\,$\mu$m eclipse depth for HAT-P-8b
(Sec.~5.4), and recovery of a previous intractable eclipse of WASP-12b
(Sec.~5.5).

The GJ\,436 data were originally used to search for an additional
planet \citep{ballard}, and they contain no transits or
eclipses. CoRoT-2b was observed at the start of Spitzer's extended
warm mission and analyzed using the polynomial method
\citep{deming11}.  WASP-14b \citep{blecic} and WASP-8b
\citep{cubillos} were both analyzed quite recently, and used the BLISS
method \citep{stevenson12}.  WASP-8b at 3.6\,$\mu$m was a challenging
data set, which exhibited significant red noise in the normalized
light curve \citep{cubillos}.  \citet{cowan12} found that the WASP-12b
eclipse we analyze was especially difficult to fit, and they omitted
that eclipse from their results, as did \citet{stevenson14}.  We thus
challenge PLD with both variety and difficulty.  Observational
parameters for these five data sets are summarized in Table~2.

\subsection{Testing PLD With Real Data:  PLD {\it vs.} a Weighting Function}

Our initial use of PLD showed immediately that it was a powerful
technique.  We therefore worried that it might be able to re-shape the
data and produce an eclipse at any arbitrary orbital phase, even if no
real eclipse was present.  We alleviated this concern by applying PLD
to the contiguous 33-hour time series data for GJ\,436, used by
\citet{ballard} to introduce the weighting function method, and search
for transits of a possible GJ\,436c planet.  These data contain no
eclipses or transits, as \citet{ballard} discuss.

The GJ\,436 data comprise 488960 images (7640 data cubes each
containing 64 frames).  Since there are so many images with short
exposure times (0.1 seconds), binning the data for the decorrelation
is especially appropriate.  Our fit procedure (Sec.~3.3) selects bins
of 392 exposures (about 51 seconds of real time) for the
decorrelation. Our solution quickly revealed a sharp transient rise in
intensity over the first $\sim30$ to $60$ minutes of the time series.
Since this sharp increase is not adequately reproduced by the quadratic time
dependence in our Eq.(4), we used an exponential time ramp, and we
omitted the first 36 minutes of data that showed the greatest ramp effect.

In this case, there is no eclipse or transit, and the best-fit eclipse
depths are consistent with zero.  The central phase can therefore take
any value and still produce an equivalent fit.  We use the regression
solution to initialize three independent MCMC chains.  The MCMC chains
use the 392-point binning, and we find that the solution using that
binning works well when applied to the unbinned data.  We apply the
best-fit binned solution to the original unbinned data to form
residuals.  We then re-bin those residuals on various time scales, as
described in Sec.~3.3, to compare to \citet{ballard}, and to search
for red noise that may remain in the results.  We find a scatter of 65
ppm for 20-minute re-bins, shown in the top panel of Figure~4.  That
compares well to \citet{ballard}, who found 72 ppm on that time scale.

Applying our best-fit $c_i$ values and ramp parameters to the unbinned
GJ\,436 data, we re-binned the residuals on various time scales up to
$2^{16}$ exposures (about 2.5 hours), and we find that the slope of
$\log(\sigma)$ varies as $\log(N)$ with a slope of $-0.504$,
essentially identical to the $-0.5$ slope expected for Poisson noise.
The dependence of $\log(\sigma)$ on $\log(N)$ is shown in the lower
panel of Figure~4.  The unbinned residuals have a standard deviation
of $0.00548$, 3\% greater than the $0.0053$ obtained by
\citet{ballard}.  Because PLD operates on binned data, we do not fully
correct the effect of short term pointing fluctuations.  But the
effective removal of red noise is a very acceptable trade for a 3\%
increase in the short-term scatter.

We conclude that PLD passes this test in the sense that it does not
produce distortions in the data that would be mistaken for real
transits or eclipses.  We further conclude that the binned PLD
solution effectively removes noise on longer time scales, albeit at
the price of a small increase in noise on shorter time scales.

\subsection{Testing PLD with Real Data: PLD {\it vs.} Polynomial Decorrelation for CoRoT-2b}

Eclipses of CoRoT-2b were among the first data analyzed from the Warm
phase of the Spitzer mission \citep{deming11}, and \citet{hansen}
mention CoRoT-2 as one of the most significant examples of deviation
from a blackbody.  We have tested PLD on the 3.6\,$\mu$m eclipse of
CoRoT-2b reported by \citet{deming11}.  The original analysis used the
traditional polynomial method of decorrelating, with a quadratic
function in Y, and a linear function in X.  Our PLD result (not
illustrated here) differs from the polynomial solution by less than
$1\sigma$ in both the depth and phase of the eclipse. Like all of our
eclipse fits in this paper, the characteristics and results of the fit
are summarized in Table~1.  The PLD solution yields a slope of
$\log{\sigma}$ vs. $\log(N)$ of $-0.495$, statistically
indistingusihable from $-0.5$.  The slope derived from the original
solution was not reported by \citet{deming11}, but the errors on the
eclipse depth and phase for the original fit are essentially the same
as our current PLD solution.  We conclude that our PLD analysis
matches the original fit for this eclipse of CoRoT-2b, and again our
PLD fit has essentially no red noise.

\subsection{Testing PLD with Real Data: PLD {\it vs.} BLISS Mapping for WASP-14b}

Turning to eclipse data that is more recent than CoRoT-2, we
re-analyze the 3.6\,$\mu$m eclipse of WASP-14b, as originally reported
by \citet{blecic}.  We choose WASP-14b because it has a slightly
eccentric orbit, and the phase of the eclipse adds a useful dimension
to the test.  Moreover, the analysis described by \citet{blecic} uses
the well-documented and effective BLISS method \citep{stevenson12} to
correct for intra-pixel effects. 

Our PLD decorrelation of WASP-14b used a 10-frame binning (about
one-sixth of a data cube, 20 seconds of time), chosen by our PLD code
to give the best red noise removal.  \cite{blecic} found that the
choice of temporal ramp model was ambiguous, even using the Bayesian
Information Criterion. We use a linear ramp and achieve a
point-to-point scatter (standard deviation of the normalized
residuals, $SDNR=3054$\,ppm, Table~1).  Our solution improves on the
SDNR quoted by \citet{blecic}, who found values near $3310$\,ppm
(their Table~3).  Like \citet{blecic}, we omit the first 1100 of 13760
frames from the solution.  Following the regression solution of
Eq.(4), we run three MCMC chains, each having $10^6$ steps, and we
check convergence by comparing these independent chains.  Our best-fit
eclipse depth from the regression using the linear ramp is $1981\pm66$
ppm, where the error comes from fitting a Gaussian to the posterior
distribution.  We obtain essentially the same eclipse depth from the
centroid of the symmetric posterior distribution ($1968$ ppm).  All of
our PLD posterior distributions for the depth and central phase of all
the eclipses we analyze are indistinguishable from Gaussians.  Our
result ($1981\pm 66$) ppm is in excellent agreement with
\citet{blecic}, who quote an eclipse depth of $1900\pm100$ ppm.  Using
the eccentric orbit model from Table~10 of \citet{blecic}, our
best-fit central phase is $0.4833\pm0.0004$, compared to
$0.4825\pm0.0003$ from \citet{blecic}, a $2\sigma$ difference.  Both
solutions confirm an eccentric orbit, and the central phases differ by
a marginally significant amount.

Figure~5 shows our best-fit PLD eclipse for WASP-14b (top panel),
re-binned to approximately the same time resolution used for Figure~8
of \citet{blecic}.  Comparing to the overplotted (gray) points from
\citet{blecic}, the PLD fit has fewer outliers, but is otherwise very
similar.  The middle panel shows the residuals from our fit, and the
lower panel shows the standard deviation of our residuals when binned
on time scales from one frame (2 seconds) to $2^{11}$ frames (about
4100 seconds, including overhead).  The best-fit slope to the binned
standard deviations on the lower panel of Figure~5 ($-0.494$) is
statistically indistinguishable from Poisson noise ($-0.5$).

We conclude that our PLD analysis gives an eclipse depth consistent
with previous work, and again we find that it attenuates red noise.

\subsection{Testing PLD with Real Data: PLD {\it vs.} BLISS Mapping for WASP-8b}

Another challenging test of PLD is the 3.6\,$\mu$m eclipse of WASP-8b
reported by \citet{cubillos}.  Those authors utilized a BLISS
technique to analyze this eclipse, but found significant red noise
remaining after the decorrelation.  Moreover, they found an eclipse
depth that they described as `anomalously high', requiring some
tension in the astrophysics to account for it (Sec.~6 of their paper).
We have applied PLD to this eclipse, using a quadratic temporal ramp
as per the original analysis \citep{cubillos}.  Both the original, and
our re-analysis, omit some frames at the start of the time series,
which is a normal procedure for Spitzer analyses because there are
transient effects at the outset. 

This star has a possibly bound M-dwarf companion 5 arcsec distant, and
2.1 magnitudes fainter in K-band \citep{queloz, cubillos}.  The
companion lies outside of our photometry apertures, but can contribute
scattered and diffracted light.  Appealing to symmetry, we measure the
scattered and diffracted light from WASP-8 itself, by placing a
numerical aperture at the distance of the companion, but on the
opposite side from it.  We calculate a correction factor using that
fractional light contribution together with the relative brightness of
WASP-8 and the companion. We apply the dilution correction to the
eclipse depth after the decorrelation and fitting process, not to the
photometry. The correction we calculate is 2.5\%.

Our PLD fit is listed in Table~1, and shown in Figure~6. We find a
significantly cleaner fit to the eclipse with fewer outliers compared
to the original fit from \citet{cubillos} (compare black and gray
points in the top panel of Figure~6.) For WASP-8, as well as HAT-20
(discussed below), eclipses shown graphically in our Figures do not
include correction for the companion stars, but that correction is
included in the Table~1 eclipse depths.  Our code chooses 148 point
binning (1.05 minutes) for our WASP-8 solution, i.e. slightly more
than two Spitzer data cubes.  Applying that binned solution to our
unbinned photometry, we find a SDNR$=5414$\,ppm, vs. $5377$ from
\citet{cubillos}, a 0.7\% difference.  Our PLD solution yields a slope
of the binned-$\sigma$ relation of $-0.492$, showing essentially no
red noise.  Our eclipse depth is $906\pm74$\,ppm, vs. $1130\pm180$
from \citet{cubillos}, a difference that is 1.2 times their error.
\citet{cubillos} determine the precision of their eclipse depth by
accounting for correlations in their residuals using the $\beta$
parameter method of \citet{winn}.  Our error is based solely on our
MCMC posterior distribution for eclipse depth, since we find no
significant correlation in the residuals.  Our eclipse depth is
consistent with the atmospheric models for the planet shown in
Figure~8 of \citet{cubillos}, and reduces the tension with the
astrophysics of the exoplanetary atmosphere that they discuss.
Although we believe that the eclipse depth from \citet{cubillos} is
too large, we point out that their excellent error analysis
encompasses our revised eclipse depth.
 
Sec.~2 mentions that it might be possible to obtain good
photometry from the sum of the $3\times3$-pixels in Eq.(3). As the
stellar image moves, the fraction of total light encapsulated by a
$3\times3$-pixel sum will vary, because the image is moving but the
pixels are stationary.  It is reasonable to hypothesize that Eq.(4)
will correct for variable light loss, just as it corrects for the
intra-pixel sensitivity effect.  If so, it may be possible to obtain
excellent eclipse results by simply summing the pixels that contain the
star, without implementing conventional aperture photometry. Simple
sum-of-pixels photometry has distinct advantages.  It obviates all of
the issues associated with the best way to measure the stellar
centroid, and other effects such as the `pixelization' discussed by
\citet{stevenson12} become irrelevant.

Sum-of-pixels photometry is most appropriate for bright stars, where
the stellar intensity greatly exceeds the sky background.  For
relatively faint stars where background fluctuations contribute
significantly to the noise, aperture photometry remains desirable in
order to optimize the star-to-background ratio.  For that reason, we
continue to rely on aperture photometry as a primary tool in our PLD
analyses, but we here test sum-of-pixels photometry for our brightest
eclipsing system, WASP-8.  Figure~7 compares the aperture photometry
for WASP-8 (Figure~6) with an independent decorrelation based on
replacing the aperture photometry with the denominator of Eq.(3),
i.e. using the sum of a $3 \times 3$-pixel box.  The results are very
similar (Table~1); the eclipse depths differ by only 54 ppm, less than
$1\sigma$, and the Gaussian-shaped posterior distributions for eclipse
depth (Figure~7) overlap significantly.  The slopes of the
binned-sigma relations (Table~1) are both indistinguishable from
$-0.5$.  Moreover, both results for the eclipse depth are in good
agreement with the modeled spectrum shown by \citet{cubillos},
eliminating the need to invoke unusual astrophysics.

We conclude that PLD permits robust photometry of bright stars,
without the need to measure the position of the image.

\subsection{Testing PLD with Real Data: WASP-12b}

\citet{cowan12} studied the phase variation of thermal emission from
the very hot planet WASP-12b, and their Spitzer data contained two
eclipses of the planet at 3.6\,$\mu$m.  The first of these eclipses
exhibited `highly correlated residuals' after their polynomial and
weighting function decorrelations \citep{cowan12}. Their depth for
this problematic eclipse was significantly less ($0.0030$ versus
$0.0038$) than a previously analysed eclipse of this planet also at
3.6\,$\mu$m \citep{campo}. \citet{stevenson14} declined to include
this eclipse in their recent re-analysis of WASP-12b data. It
therefore makes a challenging case for our PLD analysis.

Our best-fit solution (Figure~8) for this eclipse yields a depth of
$0.00363\pm0.00018$, consistent (at $1\sigma$) with the second eclipse
in the data analyzed by \citet{cowan12} ($0.0038\pm0.0004$).  We also
agree with other 3.6\,$\mu$m eclipses analyzed by \citet{campo}
($0.00379\pm0.00013$), and approximately with \citet{stevenson14}
($0.0041\pm0.0002$, $0.0038\pm0.0002$, and $0.0036\pm0.0002$). Our
best-fit eclipse phase using the ephemeris from \citet{chan} is
consistent with a circular orbit \citep{campo}.  Although the eclipse
depths quoted above do not include correction for the dilution by the
companion star, that correction is included in Table~1.  Since our
average photometric aperture is close to the 3.0 pixels used by
\citet{stevenson14}, we adopt their dilution correction factor
($1.1149$).  Our corrected eclipse depth ($4051\pm202$\,ppm) agrees
well with the corrected average eclipse depth (3 eclipses) from
\citet{stevenson14} ($4210\pm110$\,ppm), and one eclipse from
\citet{campo} ($3790\pm130$\,ppm).

Our result for this eclipse is compared to \citet{cowan12} in
Figure~8.  Some important differences in method are that
\citet{cowan12} were fitting an entire orbit of data, and they used a
planetary phase function, but no instrumental temporal ramp.  We fit
to only the data between phases 0.4 and 0.6, and we use a linear
temporal ramp in Eq.(4). Those differences alone will tend to give us
better results for the eclipse, but our intent is primarily to
demonstrate a successful PLD analysis of this eclipse, and only
secondarily to compare to \citet{cowan12}.  The fit from
\citet{cowan12} exhibits a slope in the residuals during the eclipse,
when the planet is not contributing. Figure~8 shows that we find
significantly less slope, as can be seen particularly just prior to
egress.  We also find less scatter, and very little correlation in the
residuals.  Our slope for the binned residuals (bottom panel of
Figure~8) is $-0.470$.  We conclude that PLD can successfully fit this
difficult eclipse data set.

\section{Observations of HAT-P-20b, and Initial Data Processing}

We now turn to HAT-P-20b \citep{bakos}, and apply PLD to this
moderately-irradiated giant exoplanet.  We here analyze four eclipses that
have not previously been published.

We observed two eclipses of HAT-P-20b in each Warm Spitzer band, in
program 80219 (H. Knutson, P.I.), using subarray mode. Times of the
observations are given in Table~3.  Our analysis used the BCD data
cubes, each having 64 frames of 2-second exposures.  We find and
correct discrepant pixels due to energetic particle hits or other
transient effects using a median filter applied to each pixel as a
function of time.  We construct a 5-pixel running median of each
pixel's intensity within a given $32\times 32 \times 64$-pixel data
cube, and we set $4\sigma$-discrepant pixels to the median value.  We
similarly apply a $4\sigma$ median filter to the photometry and image
positions internal to each data cube.

For both HAT-P-20b and the tests using synthetic and archival data, we
subtract a background level from each frame of the 64-frame data cube,
by fitting a histogram to pixels in the four $6 \times 6 \times
64$-pixel spatial corners of each data cube. We use this procedure
to minimize background contribution from HAT-P-20 itself and from a
companion star \citep{bakos}.  After the background subtraction, we
locate the center of the HAT-P-20 stellar image using both a 2-D
Gaussian fit, and also a center-of-light calculation.  We measure the
flux using both constant-radius and variable-radius apertures, as
described in Sec.~4.2.

HAT-P-20's companion star is physically bound \citep{ngo}, about one
magnitude fainter, and 6.9 arc-sec (5.7 pixels) distant.  As for the
WASP-8 case, we estimated the (small) contribution of diffracted light
from the companion by measuring the flux from HAT-P-20 6.9 arc-sec in
the opposite direction from the companion, using the same photometric
apertures that we adopted for HAT-P-20.  After adjusting for the
brightness of the companion relative to HAT-P-20, we find that the
depth of the HAT-P-20 eclipses are diluted by 0.63\% and 1.56\% at
3.6- and 4.5\,$\mu$m respectively, and we applied this correction to
our results after the decorrelating and fitting process.  We also used
photometry of the companion star as a check on our results for the
HAT-P-20 eclipses, as described in Sec.~7.2.

\section{Eclipses of HAT-P-20b}

Table~1 lists the parameters of the best-fit for each eclipse of
HAT-P-20. As for previous eclipses, all of our PLD solutions for
HAT-20 have a slope of $\log{\sigma}$ vs. $\log{N}$ close to the
Poisson value of $-0.5$, but we do not illustrate the $\sigma(N)$
relations in these cases.

Figure~9 illustrates the unbinned {\it vs.} binned aspect of our fits,
using the second HAT-20 eclipse at 3.6\,$\mu$m, that has the most
binning (maximum binning facilitates seeing the difference).  The top
panel shows the unbinned photometry overlaid point-by-point with the
best fit calculated using the $c_i$ coefficients from the binned fit.
The middle panel shows the binned data and the binned fit, using a
48-exposure binning selected by our fit procedure (Sec.~3.3).  The
lower panel shows the residuals for the unbinned case (data minus
fit), showing the close resemblence of the residuals to white noise.

\subsection{3.6\,$\mu$m Eclipses}

Figure~10 shows the two eclipses at 3.6\,$\mu$m.  Our analysis code
selects a wide variety of bin sizes when doing the HAT-P-20 fits (1, 2, 
32, and 48 exposures). Consquently, for Figure~10 we re-bin the
photometry so that 50 points span the data.  The lower panel of
Figure~10 shows the posterior distributions from the MCMC chains.  One
chain per eclipse is illustrated, but we used three independent chains
of $10^6$ steps for each eclipse and their distributions were closely
identical.  The two eclipses at this wavelength differ in their
retrieved depth, but the difference is only 246 ppm (see Table~1).
Since these are independent events, the error on the difference in
eclipse depths equals the quadrature sum of the errors on the
individual eclipses, which is 163\,ppm.  So the difference in the two
eclipse depths at 3.6\,$\mu$m is $1.5\sigma$, consistent with random
noise.

We found one anomaly in our PLD solutions.  The eclipse depth for the
second eclipse at 3.6\,$\mu$m is degenerate with the purely temporal
(baseline) terms in Eq.(4). We use a quadratic temporal ramp in the
solution, but the degeneracy remains if we use only a linear ramp (and
the fit is worse). We also explored using an exponential ramp in
Eq.(4) for this eclipse, but we find that it does not produce an
acceptable fit.  Examining the fit closely, we found that the data
required a `U'-shaped baseline, and an exponential cannot produce that
shape. The 'U'-shape is evident on the middle panel of Figure~9.  We
therefore adopt a quadratic temporal ramp, and we tolerate the
degeneracy because it is included in the error derived from the
posterior distributions - note the broader distribution for the second
3.6\,$\mu$m eclipse on Figure~10.  Figure~11 shows the MCMC chain
values for the baseline coefficients and the pixel ($c_i$)
coefficients for this eclipse.  The degeneracy is obvious from the
correlations shown in panels on the top right, giving the linear and
quadratic coefficients of time. Note also that the linear coefficient
is correlated with the quadratic coefficient, since the fit can
compensate for less or more baseline curvature by varying the baseline
slope.  None of the $c_i$ coefficients exhibit any correlation with
the eclipse depth.  Nor do we find correlations between the $c_i$ and
the eclipse depths for any data set we have analyzed.

\subsection{4.5\,$\mu$m Eclipses}

Figure~12 shows the two eclipses at 4.5\,$\mu$m. For visual clarity,
they are binned to 50 (first eclipse) and 40 points (second eclipse
having less data).  In this case, the difference in eclipse depths
(625\,ppm) is about four times the error of the difference
(154\,ppm). This $4\sigma$ difference is not a statistical
fluctuation, given that the posterior distributions are closely
Gaussian. Either the errors are underestimated, or the planet is
variable.  We considered possible variable dilution by scattered light
from the companion star.  Because the companion is 6.8 arc-sec distant
from HAT-P-20, its scattered light contribution is only about 1.5\% of
HAT-P-20, and our measurements show that it does not vary sufficiently
to significantly affect the relative eclipse depths at either 3.6 or
4.5\,$\mu$m. Moreover, we find no degeneracies or any other anomalies
in our PLD fits at this wavelength, in contrast with 3.6\,$\mu$m -
where the eclipse depths are in good agreement in spite of the
degeneracy discussed above.

If the errors are underestimated, the most likely reason is that the
results depend on features of the data or decorrelation process that
are not included in the variations probed by the Markov chains.  One
such possibility is the choice of pixels used in the PLD
decorrelation.  All of our fits listed in Table~1 use 9 pixels,
usually in a $3 \times 3$-box centered on the star.  Since the corner
pixels contain the least flux, it is arguably possible that they are
unnecessary to the fit, and might even be perturbing it in an
undesirable way.

To explore the robustness of the PLD fits, we re-fit both eclipses
with the corner pixels omitted from the PLD solution.  The posterior
distributions for the no-corner fits are plotted with dashed lines on
the lower panel of Figure~12.  They are shifted slightly with respect
to the $3 \times 3$-pixel solutions, but still indicate different
depths for the two eclipses.  Also, the central phases for all of the
HAT-P-20b eclipses (Table~1) are very consistent.  We conducted additional
checks such as forcing our code to use the same parameters (binning,
centroiding, aperture type and size) for both eclipses, and the
difference between the two eclipses persists.  We also implemented a
conventional polynomial decorrelation, by replacing the
$\hat{P}_{i}^{t}$ values in Eq.(4) with image centroid coordinates
($X$, $X^2$, $Y$, and $Y^2$).  Those posterior distribution are shown
as dotted lines on Figure~12, and are in good agreement with the PLD
results.

In principle, if our values for the average brightness of HAT-P-20's
host star at each eclipse were in error by a large amount, the
resultant incorrect normalization factors could lead to large errors
in the eclipse depths.  We checked this by comparing HAT-P-20 to the
companion star.  We find the average brightness of HAT-P-20 decreased
by 1.6\% from the first to the second eclipse at 4.5\,$\mu$m, and the
companion decreased by 1.9\%.  At 3.6\,$\mu$m, HAT-P-20 decreased by
7.8\%, versus a 1.6\% increase for the companion.  Although the
relatively large variation of HAT-P-20's absolute brightness at
3.6\,$\mu$m is puzzling, it is not large enough to affect the eclipse,
and the two 3.6\,$\mu$m eclipse depths are consistent within the
errors, as discussed in Sec.~7.1.  Variation in the absolute
brightness of HAT-P-20 at 4.5\,$\mu$m is consistent with the variation
seen in the brightness of the optical companion, so there is no reason
to attribute our result to errors in normalizing the photoemtry.

We also decorrelated the photometry of the companion star using the
same PLD code as for HAT-P-20.  We solve for the depth of an `eclipse'
in the companion data, constraining it to have the same orbital phase
as observed for HAT-P-20's eclipse (Table~4), and using a simple
linear ramp in time.  These decorrelated results all show flat time
series, with a per-exposure scatter that exceeds the photon noise by
an average of 29\% and 13\% at 3.6- and 4.5\,$\mu$m, respectively.
The slopes of the binned-$\sigma$ relations were better than -0.48 in
all four cases, and the derived `eclipse' depths were consistent with
zero.  Those depths were (for the same order as Table~3):
$+46\pm99$\,ppm, $-122\pm78$\,ppm, $+68\pm118$\,ppm, and
$-181\pm113$\,ppm.

We conclude that the difference in HAT-P-20's 4.5\,$\mu$m eclipse
depths is not due to the PLD analyses.  In order to infer the average
atmospheric properties of HAT-P-20b, we calculate the average eclipse
depth in each band, weighting each eclipse by the inverse of its
variance.  Those average values are listed in Table~4.

\section{Implications for the Atmosphere of HAT-P-20b}

Figure~13 shows our results for eclipse depths of HAT-P-20b in
comparison to the contrast from a best-fit blackbody temperature of
$1134\pm29$\,K.  We estimated the error for that best-fitting
blackbody by increasing the observed error at 4.5\,$\mu$m to allow for
the discordant eclipse depths at that wavelength.  The best-fit
blackbody temperature is essentially identical to the $T=1157$\,K that
would prevail if HAT-P-20b absorbs stellar energy with zero albedo,
and re-radiates uniformly over the star-facing hemisphere.
\citet{cowan11} studied the statistics of heat re-distribution using
secondary eclipse data, and \citet{perez} studied heat re-distribution
using phase curves. Both find a tendency for the most
strongly-irradiated planets to circulate heat with the least
efficiency.  A strongly-irradiated planet will be hot, and the
radiative time constant decreases strongly with temperature.  A short
radiative time constant in turn implies that the planet re-radiates
incident stellar energy before hydrodynamics can advect it to the
anti-stellar hemisphere \citep{showman, cowan11, perez}.  HAT-P-20b is
irradiated at only a modest level (equilibrium temperature 970K for
2-hemisphere re-radiation), but the high density and probable high
metallicity of the planet should produce higher atmospheric
opacity. We suggest that high opacity may sufficiently compensate for
less irradiation, keeping the radiative time scale short compared to
advection.  \citet{lewis10} studied the day-night flux difference for
GJ\,436b using a numerical hydrodynamic model, and found that that the
difference does increase with metallicity, but only by $\sim 30\%$,
less than needed to account for HAT-P-20b.  However, HAT-P-20b is
hotter than GJ\,436b, and the effect of metallicity should be larger
at higher temperature. Also, the atmosphere of HAT-P-20b may contain
abundant absorbing clouds because the temperature is below the
condensation point for many compounds, and the metallicity may be
high. In that case, cloud absorpion could further increase the
day-night temperature difference.

Figure~13 includes two solar metallicity model atmospheres
\citep{burrows06, burrows07, fortney05, fortney06a, fortney06b,
  fortney08} that have minimal heat re-distribution.  These models
have relatively strong absorption in the 4.5\,$\mu$m band due to
carbon monoxide and water vapor \citep{sharp}, and they therefore
deviate from a blackbody model.  However, we find that a blackbody at
$1134$\,K is essentially a perfect fit to our average eclipse depths,
matching each value in Table~4 to better than $1\sigma$.  This
blackbody-like behavior frustrates our inital motivation to find
strong molecular absorption in a modestly-irradiated, metal-rich,
giant exoplanet.  However, our work does suggest possible variability
in the eclipse spectrum of this planet.

Assuming solar composition, the 4.5\,$\mu$m Spitzer band is formed
higher in the exoplanetary atmosphere than is the 3.6\,$\mu$m band.
If conditions in the atmosphere vary strongly with time, then we
expect the greatest variability at the highest altitude, because low
density regions are more easily perturbed than high density
regions. Two mechanisms can translate atmospheric variability to the
emergent spectrum: patchy clouds, and hot spots at any altitude
\citep{morley}.  Hot spots at high altitude are qualitatively
consistent with our result of divergent eclipse depths at 4.5\,$\mu$m,
but the requisite amplitude seems unrealistic.  Brown dwarfs are often
found to exhibit variability due to rotational modulation, but
HAT-P-20b would have to exhibit a much greater amplitude of
variability than do brown dwarfs. The large amplitude of apparent
variability that we observe is difficult to reconcile with our
expectations for hot Jupiter atmospheres.

An arguably more plausible explanation for eclipse depth variability
is circum-planetary thermal or fluorescent emission in the fundamental
band of CO, due to planetary mass loss or ongoing accretion.  The CO
band falls in Spitzer's 4.5\,$\mu$m bandpass, and was considered as
producing an anomalous eclipse of CoRoT-2b by \citet{deming11}.
HAT-20b is likely to be a high-metallicity planet, and the CO
abundance will increase approximately as the square of the
metallicity, so large CO column densities are plausible. Moreover, if
circum-planetary emission contaminates the 4.5\,$\mu$m eclipse depth,
then our inferred temperature for the planet will be biased too high,
and the efficiency of longitudinal heat transfer could be higher,
making that aspect of the observations less puzzling.

Claiming a high degree of variability in eclipse depth requires
strong evidence, and two eclipses - no matter how thoroughly they are
analyzed - are insufficient to conclude that this planet is variable.
However, our results {\it are} sufficient to warrant further eclipse
monitoring of HAT-P-20b.  Our working hypothesis is that
circum-planetary emission in the fundamental CO band may be
important. Monitoring of the 4.5\,$\mu$m eclipse depth by Spitzer can
establish whether there is real photometric variability, but
spectroscopic observations using JWST will be necessary to detect
possible CO emission.

\section{Implications for the Orbit of HAT-P-20b}

The central phase of all four eclipses we observe is consistently
later than the $0.5$ value for a circular orbit (Tables~1 \& 4).  We
first ask whether this could be due to the accumulated uncertainty in
the orbital period.  We use the most precise available orbital period
from \citet{granata}, but even the discovery-era period error given by
\citet{bakos} ($4.0\times 10^{-6}$ days) is already an order of
magnitude too small to account for the phase shifts we observe.  Using
the ephemeris from \citet{granata}, we calculate the average eclipse
phase at each wavelength, and the grand average for all four
eclipses. The results are listed in Table~4; we find a grand average
orbital phase of $0.50843\pm0.00041$, and the uncertainty in the
ephemeris from \citet{granata} contributes negligibly.  The light
travel time across the orbit is $36$ seconds, hence the eclipse for a
circular orbit would occur at phase $0.50014$, and our measured phase
corresponds to $e\cos(\omega) = 0.0130\pm 0.0006$.  Radial velocity
observations of this system were analyzed by \citet{knutson14}, who
derived $e\cos(\omega) = 0.013^{+0.0023}_{-0.0025}$, closely
consistent with our secondary eclipse timings.  To close the loop on
this system, we have derived new orbital parameters using a joint MCMC
fit of the RV and secondary eclipse timings, as described by
\citet{knutson14}.  The results from this fit are given in Table~5,
and illustrated in Figure~14.  The priors used in the fit are the RV
observations reported by \citet{knutson14}, the transit ephemeris from
\citet{granata}, and the secondary eclipse timings from Table~4.  Our
result of $e\cos{\omega}= 0.01352^{+0.00054}_{-0.00057}$ establishes
the small eccentricity of the orbit to high statistical confidence.
Given the existence of a bound stellar companion, HAT-P-20b is another
excellent candidate for orbital evolution via Kozai migration
\citep{fabrycky}, or other three-body mechanism.

\section{Summary}

In this paper we have introduced a new method for correcting the
intra-pixel effect in Spitzer photometry at 3.6- and 4.5\,$\mu$m, that
we call pixel-level decorrelation (PLD).  PLD differs fundamentally
from all previous methods because it removes the effect of positional
jitter without explicitly using the position of the stellar image.  We
argued the conceptual advantages of PLD (Sec.~2), and we have tested
it using both synthetic (Sec.~4) and real data (Sec.~5).  We point out
that all methods to decorrelate Spitzer photometry at these
wavelengths are subject to the mathematical reality that the solution
is a function of the time scale (i.e., degree of data binning) because
both the dependent and independent variables contain random error
(Sec.~3.1). Moreover, there are physical reasons to apply PLD to
binned data, discussed in Sec.~3.2.  Our fitting procedure finds the
best fit to Spitzer data by considering a range of time scales,
yielding a broad bandwidth solution having minimal red noise
(Sec.~3.3).

Our tests of PLD exploited a new capability to generate synthetic
Spitzer data, developed at the Spitzer Science Center.  These tests
began with synthetic data having no planet and no photon noise,
thereby isolating the intra-pixel detector effect (Sec.~4.2).  We
tested PLD using synthetic data for WASP-52b, and we recovered the
correct transit and eclipse depth to within $1\sigma$ (Sec.~4.3).  We
also recovered the phase curve amplitude of WASP-52, but our PLD
result was off by $3\sigma$. The large image motion that accumulates
over the time scale of a phase curve measurement is beyond the range
of applicability for our current version of PLD, so it is not yet
applicable to phase curve measurements.  However, PLD is very robust
for transits and eclipses.  We tested PLD on five real systems.  In
cases where there is no reason to doubt previous measurements, our PLD
result agrees with published results.  These cases include GJ\,436b
(Sec.~5.1), CoRoT-2b (Sec.~5.2), and WASP-14b (Sec.~5.3). In two
systems (WASP-8b, Sec.~5.4, and WASP-12b, Sec.~5.5) our PLD eclipse
depths are more astrophysically plausible than the original published
results, and have smaller random errors.  For example, our error for
WASP-8b at 3.6\,$\mu$m improves on the result from \citet{cubillos} by
more than a factor of two, and our eclipse depth agrees well with the
same modeled spectrum they used to account for eclipses in other
Spitzer bands.

We apply our PLD analysis to two eclipses of HAT-P-20b at each Spitzer
wavelength (Secs.~6~and~7). We find that the average spectrum of the
planet is very close to a blackbody at $1134\pm29$K, indicating a low
albedo and little if any longitudinal re-distribution of stellar
heating (Sec.~8). Our results at 4.5\,$\mu$m (Sec.~7.2) yield two
eclipse depths that differ by $4\sigma$.  Although two eclipses are
not enough to conclude that the planet's spectrum is variable, we do
conclude that there is justification to monitor the eclipse depth at
4.5\,$\mu$m using Spitzer, and to search for circumplanetary emission
in the 1-0 fundamental CO band using JWST.  All four of our measured
eclipses occur at a phase later than 0.5, indicating a slightly
elliptical orbit.  A joint MCMC fit of our eclipse times with RV
observations and and the transit time yields $e\cos{\omega}= 0.01352
^{+0.00054}_{-0.00057}$, and establishes the small eccentricity of the
orbit to high statistical confidence (Sec.~9).  Given the existence of
a physically bound companion star, HAT-P-20b is another candidate for
orbital evolution via Kozai migration, or other 3-body process.

\section{Acknowledgements}
We thank Jasmina Blecic, Patricio Cubillos, and Joseph Harrington for
sending us digital version of their results, used in Figures 5-7, and
we thank Julie Moses for comments on this paper.  This work is based
on observations made with the Spitzer Space Telescope, which is
operated by the Jet Propulsion Laboratory, California Institute of
Technology, under a contract with NASA.
%%%%%%%%%REFERENCES%%%%%%%%%

%%%%%%%%%FIGURES%%%%%%%%
\clearpage

\begin{figure}
%  \epsscale{0.6} \plotone{fig1.pdf}
%  \plotfiddle{fig1.pdf}{0.5in}{180}{346}{432}{0}{0}
\includegraphics[scale=0.6, angle=180]{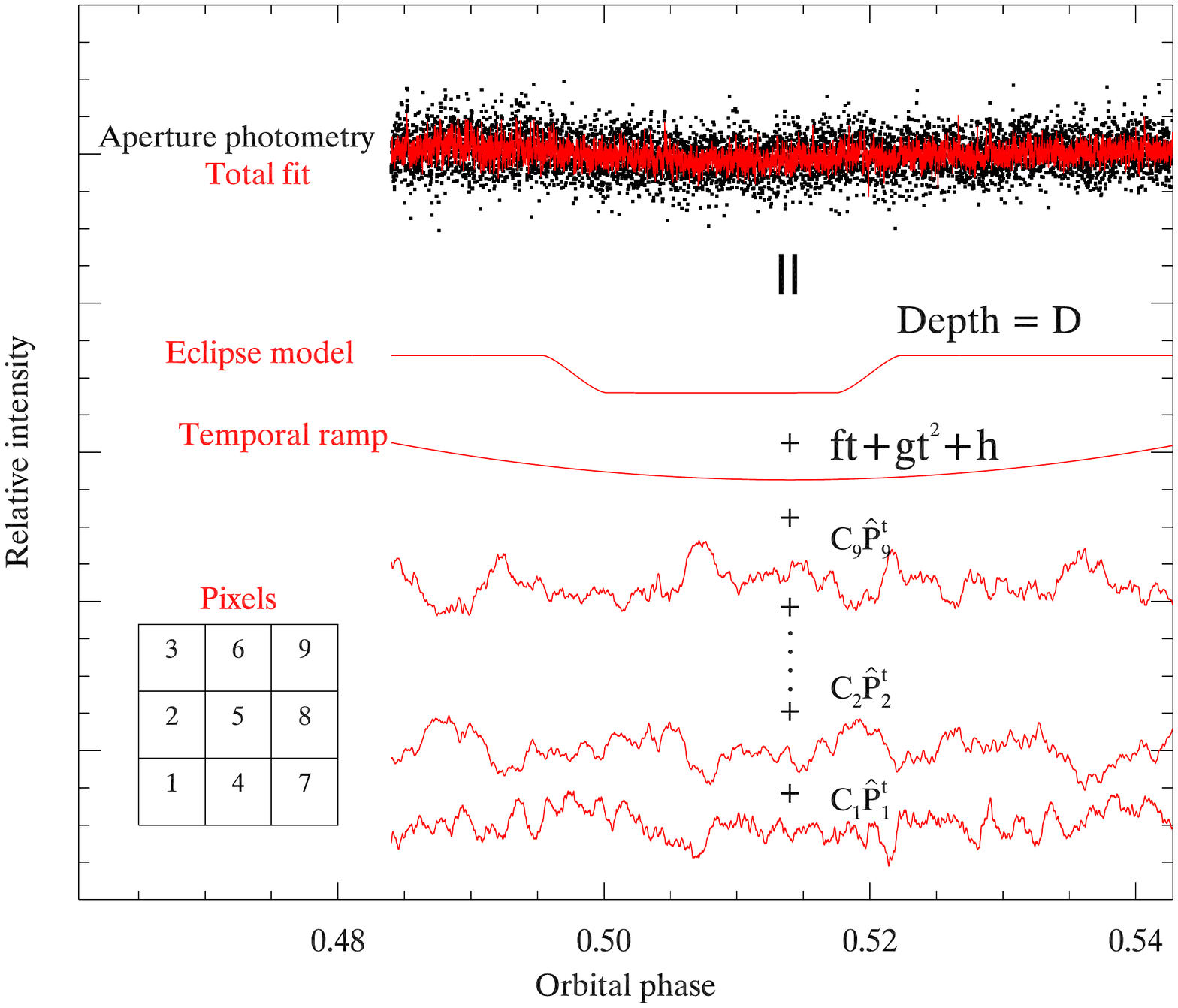}
\vspace{-0.5in} 
\caption{Graphical illustration of how PLD works, for the 2nd eclipse
  of HAT-P-20b at 3.6\,$\mu$m (see Table~3, and Figure~10). The time
  series for the relative values of the normalized pixels
  ($\hat{P}_{i}^{t}$, see Eq.~3), are each multiplied by the $c_i$
  coefficients, and added to the eclipse model and the temporal ramp
  ($t=$\,time, or orbital phase) to produce the total fit, shown in
  red overlying the aperture photometry at the top.  (These are the
  actual $\hat{P}_{i}^{t}$ time series used in the solution, but for
  clarity we here exaggerate the curvature of the temporal ramp.)  The
  inset at the lower left shows the pixel designations.  In this case
  nine pixels are used, in a $3 \times 3$ spatial arrangement.
  Nothing limits PLD to using nine pixels.  The number and spatial
  arrangement of the pixels that are actually used is determined by
  the distribution of intensity in the stellar image.}
\label{Fig1}
\end{figure}

\clearpage
\begin{figure}
\epsscale{0.9}
\vspace{-1.0in}
\plotone{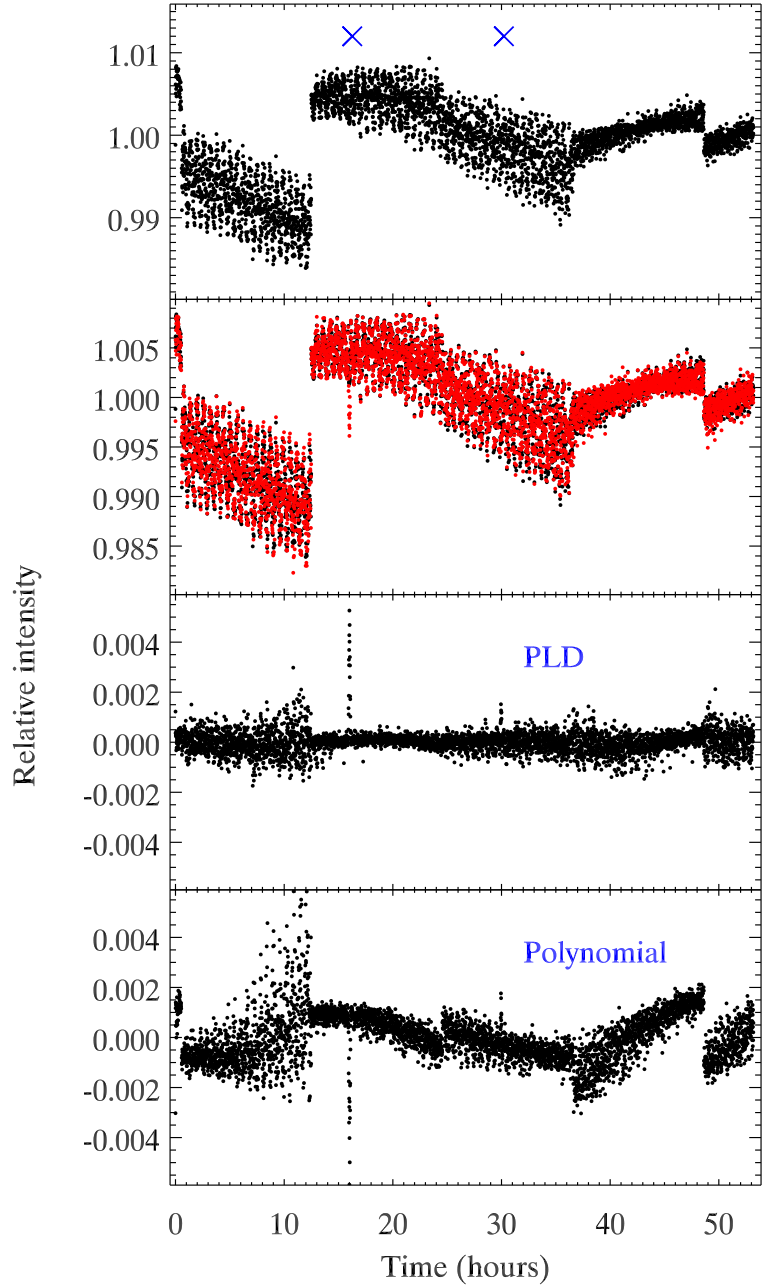}
\vspace{-0.5in} 
\caption{ Photometry of synthetic 3.6\,$\mu$m data for WASP-52, with
  the photon noise turned off and the planet removed.  All of the
  fluctuations in the top panel are due to the interaction of the
  telescope PSF with the spatial structure of the detector.  The blue
  Xs mark the times where PSF fluctuations were inserted into
  the data. The second panel overlays the PLD fit from Eq.(4), and
  the two lowest panels show the residuals for the PLD fit and also
  for a fit using a second order polynomial in both X and Y.  Note the
  different ordinate scales, especially for for lower panels.}
\label{Fig2}
\end{figure}

\begin{figure}
\vspace{-3.0in}
\plotone{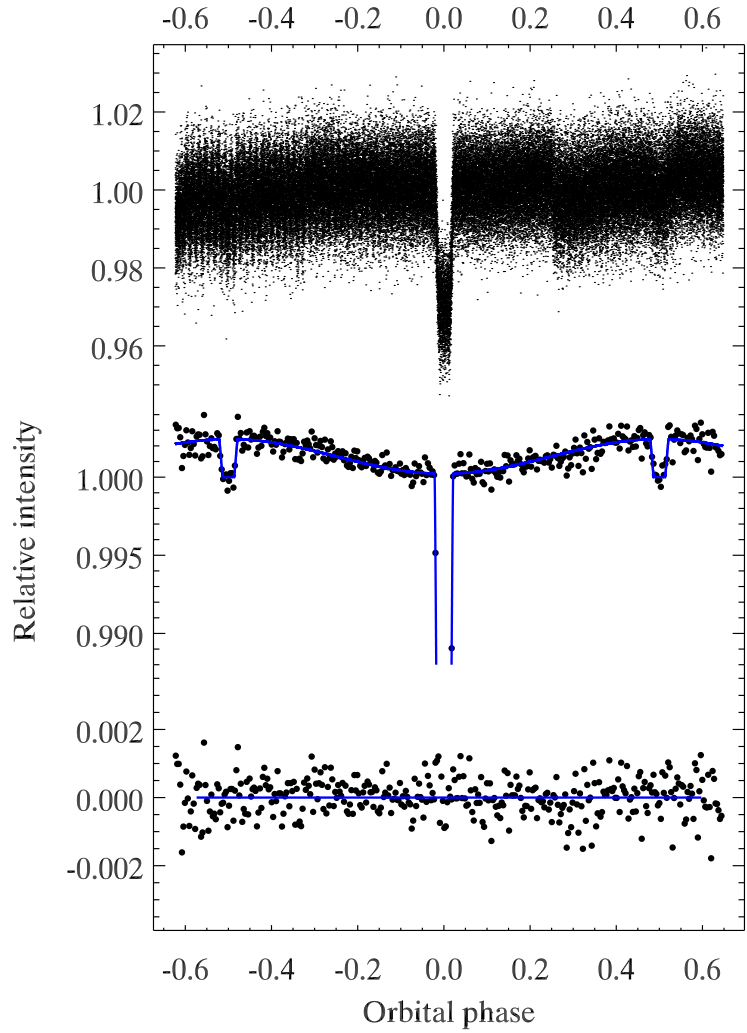}
%  \vspace{-0.5in} 
\caption{Fitting to photometry of synthetic 4.5\,$\mu$m data for
  WASP-52b, with photon and detector read noise included.  One transit
  (off scale) and two secondary eclipses of the planet are present in
  these synthetic data, as well as a phase curve modulation in
  brightness.  These are an updated version of the data used for the
  IRAC Data Challenge Workshop (see text).  Results from this fit are
  compared to the true values input to the simulation in Table~1.}
\label{Fig3}
\end{figure}

\clearpage
\begin{figure}
% \epsscale{0.5}
\vspace{-2.0in}
\plotone{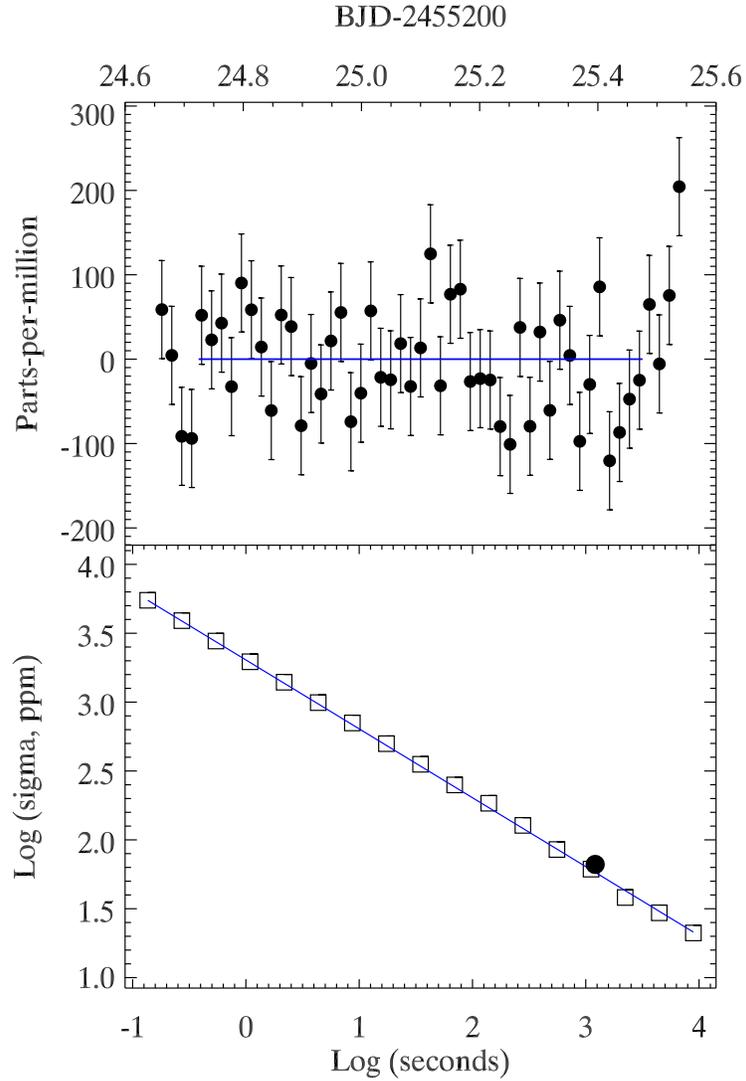}
\vspace{-0.5in}
\caption{PLD results for the 4.5\,$\mu$m time series analyzed
  originally by \citet{ballard}.  The top panel shows the residuals
  from our PLD solution, binned to a time resolution of approximately
  20 minutes (compare to Figure~6 of \citealp{ballard}) The bottom
  panel shows the standard deviation of the residual from our PLD fit,
  binned on various time scales (open points), including the 20-minute
  time scale used for the upper panel (solid point). The line is
  not a fit to these points; it's the theoretical relation that
  extrapolates the single-frame precision to larger bin sizes using a
  slope of exactly $-0.5$. }
\label{Fig4}
\end{figure}

\clearpage
\begin{figure}
%  \epsscale{0.4}
\vspace{-2.0in}
\plotone{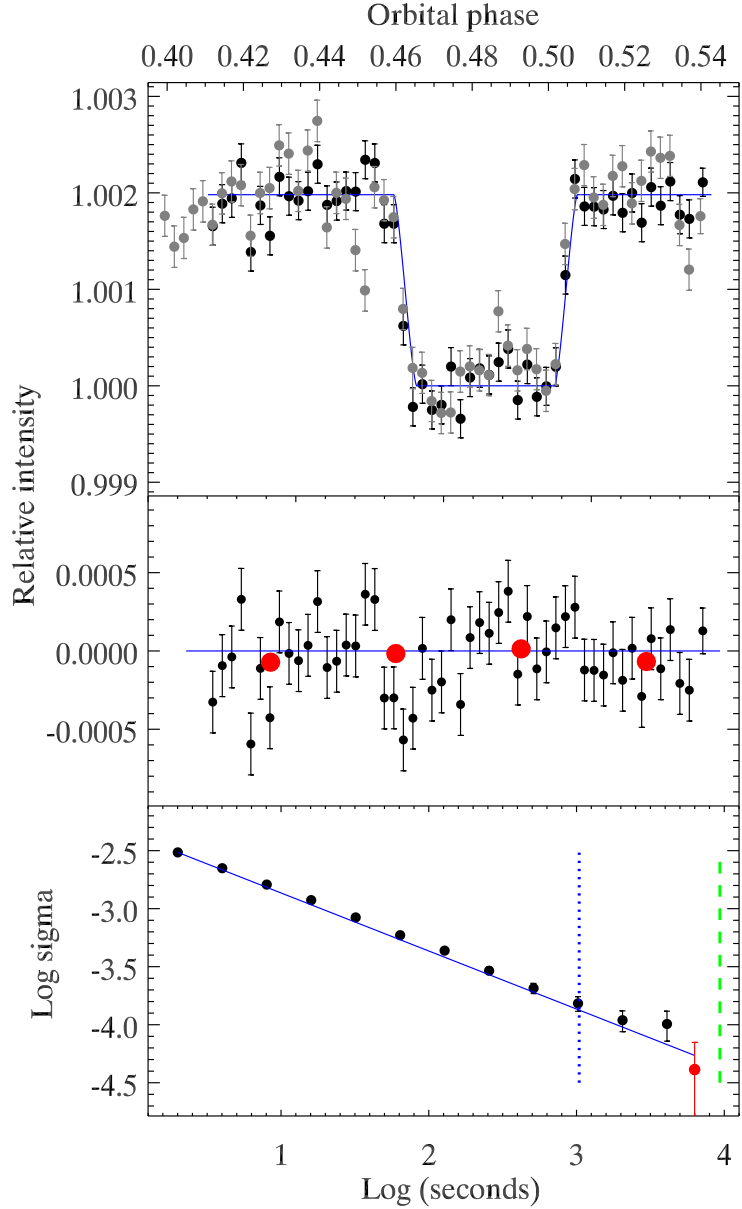}
\vspace{-0.5in}
\caption{Eclipse of WASP-14b analyzed using PLD, for comparison to
  \citet{blecic}.  The top panel shows the eclipse curve for binned
  data, using a bin size the same as \citet{blecic}. (The results
  published by \citet{blecic} are overplotted in gray.) The data and
  eclipse curve are normalized to unity in eclipse (star alone
  contributing). The middle panel shows the residuals from our fit
  (solid points with error bars), as well as a much coarser binning to
  illustrate the stability of the fit (red points).  The bottom panel
  shows the standard deviation of the residuals, at various bin sizes,
  including the bins used for the top panel and the red points in the
  middle panel.  The solid blue line is not a fit to these points;
  it's the theoretical relation that extrapolates the single-frame
  precision to larger bin sizes using a slope of exactly $-0.5$.  The
  dotted blue line marks the time scale of ingress and the dashed
  green line marks the in-eclipse time scale.}
\label{Fig5}
\end{figure}

\begin{figure}
% \epsscale{0.5}
\vspace{-2.0in}
\plotone{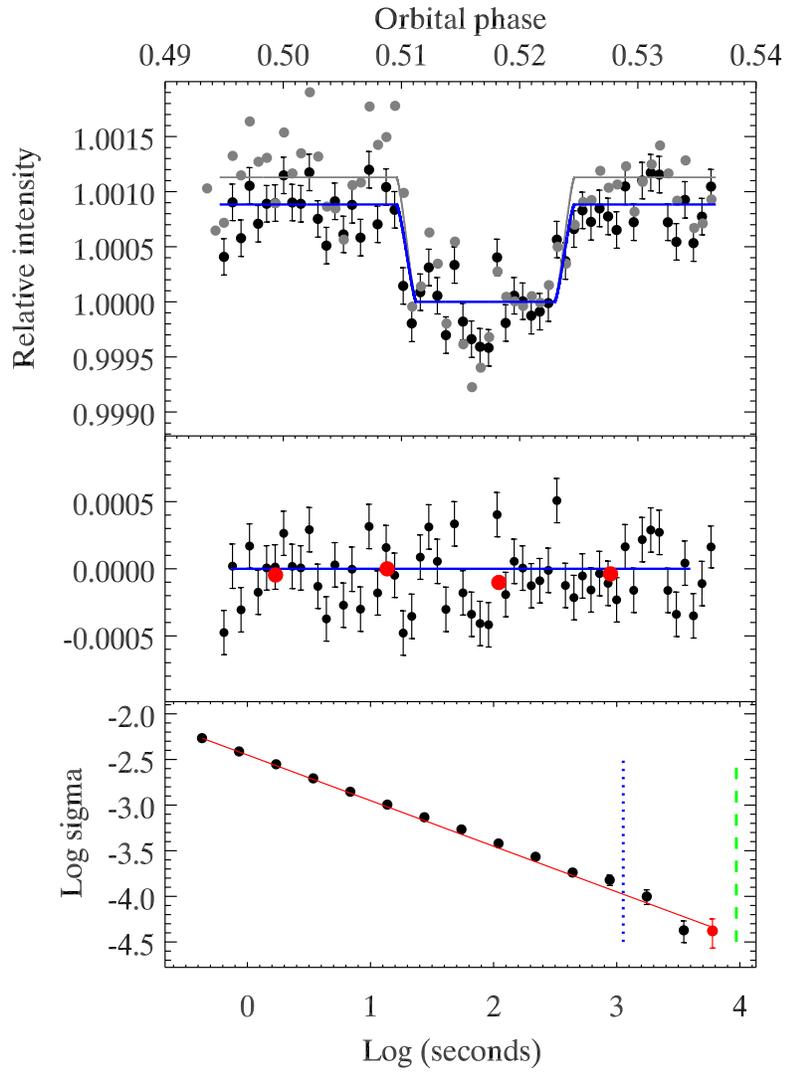}
\vspace{-0.7in}
\caption{Eclipse of WASP-8b analyzed using PLD, for comparison to
  \citet{cubillos}.  The panels are the same as for Figure~4.}
\label{Fig6}
\end{figure}

\begin{figure}
%  \epsscale{0.5}
\vspace{-2.5in}
\plotone{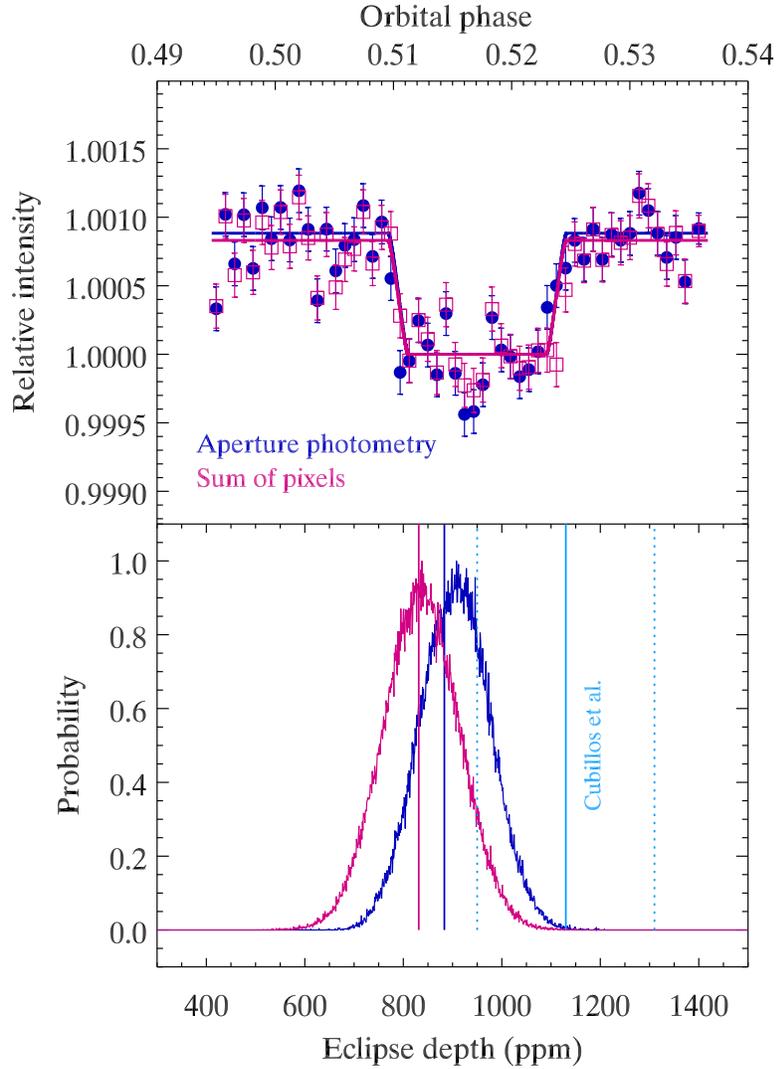}
\vspace{-0.5in}
\caption{Eclipse of WASP-8b analyzed using PLD, from aperture
  photometry and also from simple `sum-of-pxels' data that does not
  not require determining the centroid of the stellar image. The data
  and eclipse curve are normalized to unity in eclipse (star alone
  contributing). The lower panel shows the posterior distrubutions of
  eclipse depth, and the vertical lines indicate the minimum $\chi^2$
  solutions.  The eclipse depth from \citet{cubillos} is also
  indicated, with $1\sigma$ error limits (dashed lines).}
\label{Fig7}
\end{figure}

\begin{figure}
%  \epsscale{0.45}
\vspace{-1.0in}
\plotone{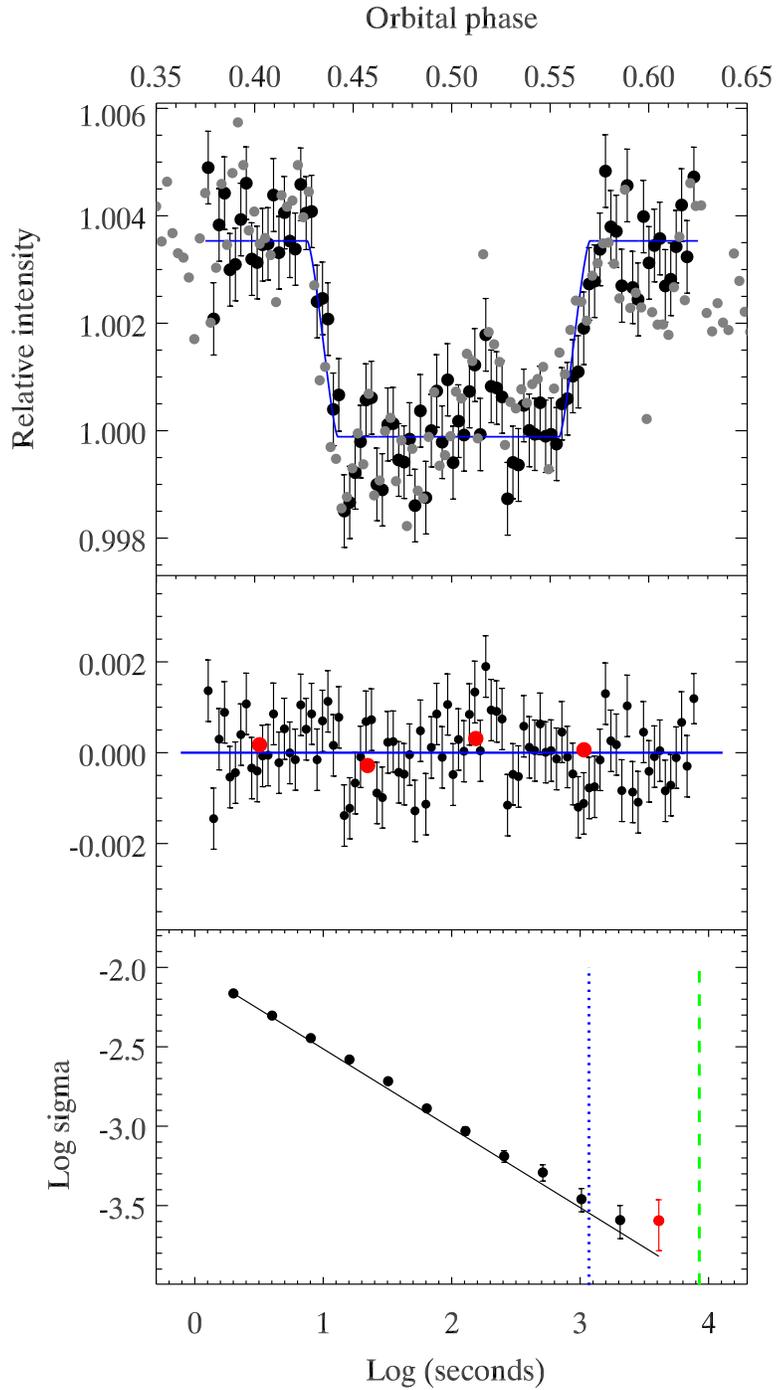}
\vspace{-0.5in}
\caption{Eclipse of WASP-12b analyzed using PLD; the panels are the same as in Figure~4. The points plotted in
gray in the top panel are from \citet{cowan12}.}
\label{Fig8}
\end{figure}

\clearpage
\begin{figure}
%  \epsscale{0.45}
\vspace{-2.0in}
\plotone{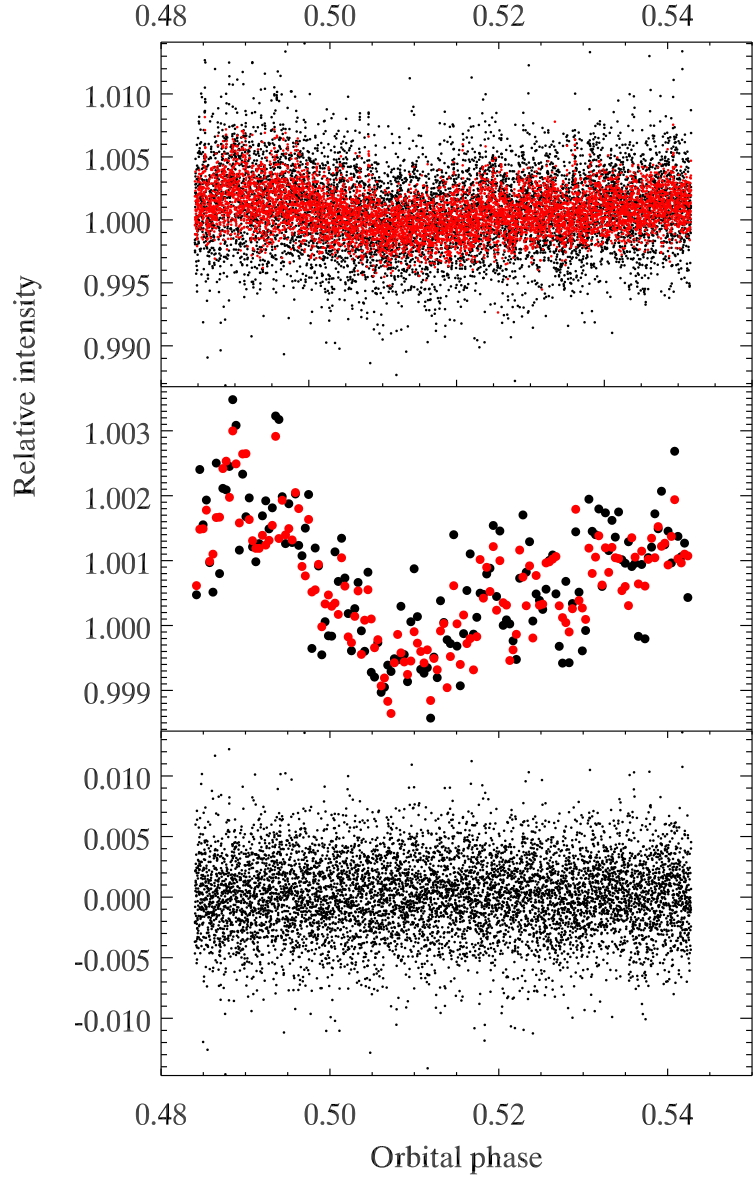}
\vspace{-0.5in}
\caption{Illustration of the unbinned {\it vs.} binned aspect of our
  fits, using the second eclipse of HAT-P-20 at 3.6\,$\mu$m.  The top
  panel shows the unbinned photometry, and the overlaid red points are
  fitted values. We include the eclipse in the fit, because we are
  here illustrating the quality of the {\it total} fit.  The middle panel
  shows the photometry binned over 48 points, overlaid by the fit (red
  points).  The $c_i$ coefficients from this fit to the binned
  photometry were used to calculate the unbinned fit values in the top
  panel.  The bottom panel shows the unbinned residuals from the top
  panel (data minus fit).  Note the white-noise-like appearance of
  these residuals.  The standard deviation of the residuals is 3150
  ppm, 22\% greater than the photon noise (2570 ppm). }
\label{Fig9}
\end{figure}

\clearpage
\begin{figure}
%  \epsscale{0.6}
\vspace{-1.0in}
\plotone{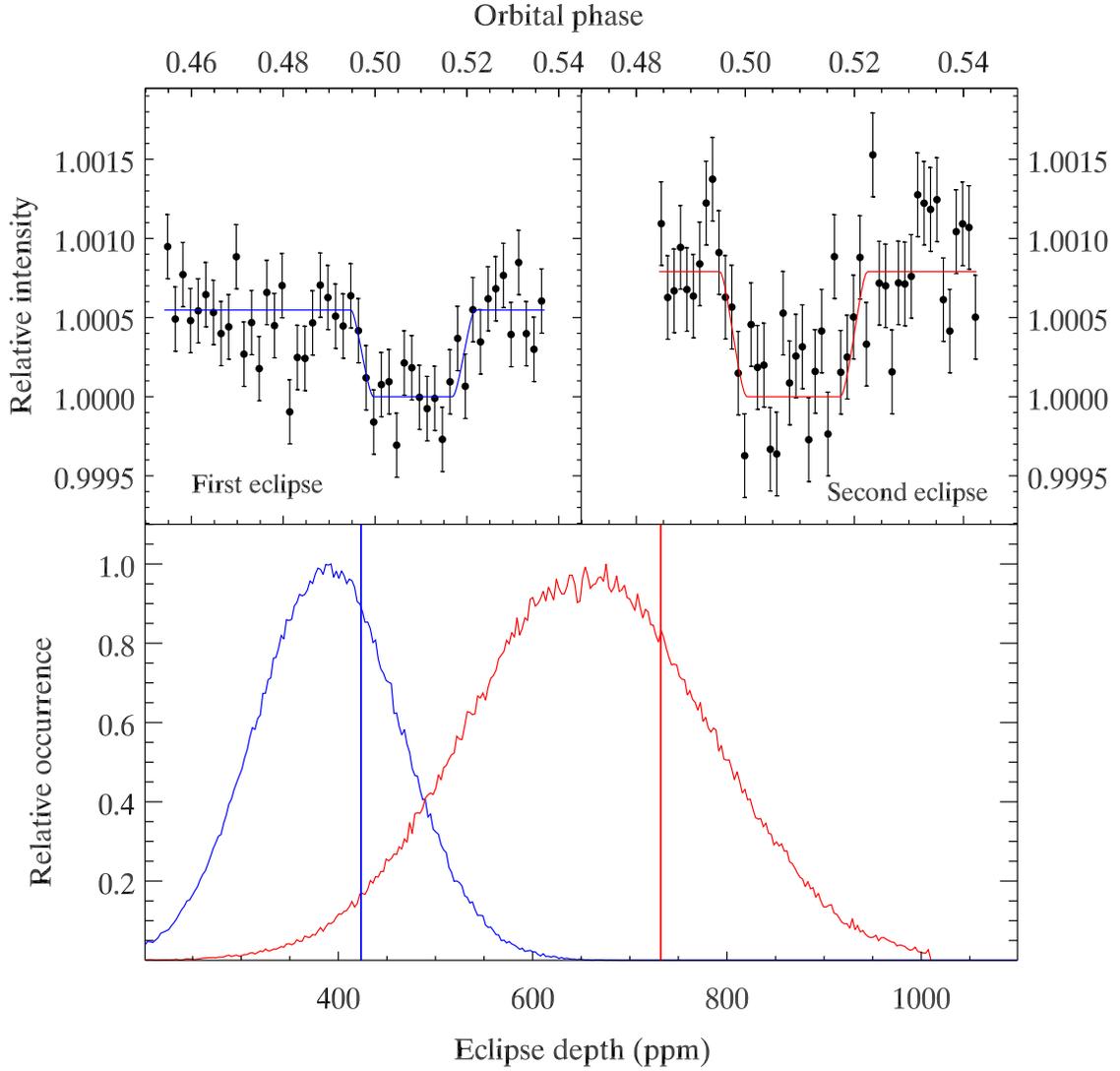}
\vspace{-1.0in}
\caption{Two PLD eclipses of HAT-P-20b at 3.6\,$\mu$m. The
  observations were binned to 50 points per data set for clarity of
  illustration. Intensity is normalized to unity in eclipse (star
  alone contributing). The lower panel shows the posterior
  distributions for eclipse depth. The vertical lines are the minimum
  $\chi^2$ values chosen using our broad bandwidth criterion
  (Sec~3.3).}
\label{Fig10}
\end{figure}

\clearpage
\begin{figure}
% \epsscale{0.7}
% \plotone{fig11.eps}
\includegraphics[scale=0.8]{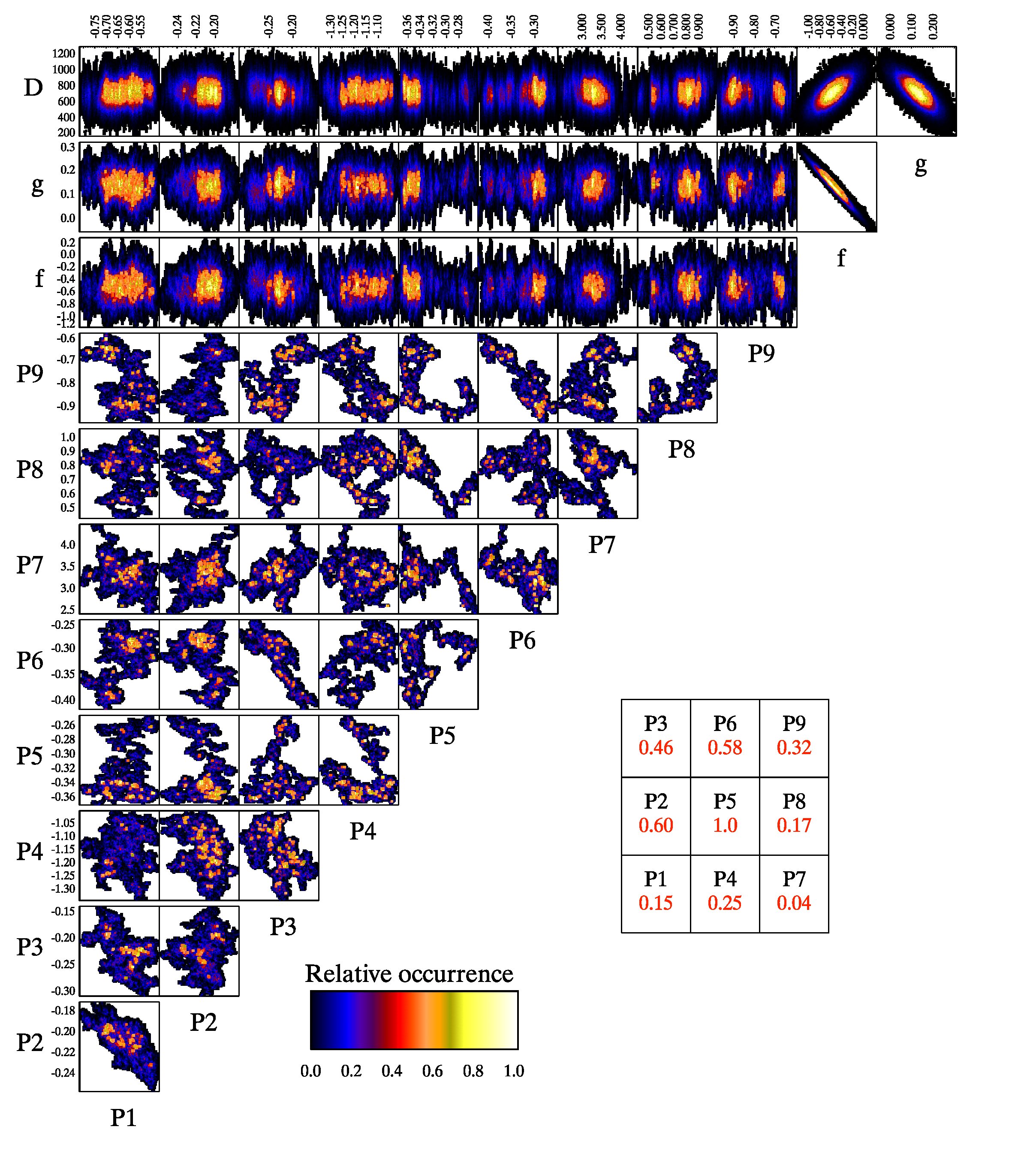}
%  \vspace{0.3in}
\caption{MCMC correlation plots for the second eclipse of HAT-P-20 at
  3.6\,$\mu$m (see Table~3, and Figure~10). The P1 through P9 panels give the
  values of the $C_i$ coefficients in Eq.~4.  The panels illustrate
  the density of all points in this $10^6$-step chain, with the color
  bar indicating the relative point density. The panels labeled $f$,
  and $g$ give the linear and quadratic coefficients of time for
  the temporal ramp, and $D$ is the eclipse depth (Eq.~4 and Fig.~1).
  The inset shows the designations of the 9 pixels and their average
  relative values (red numbers).}
\label{Fig11}
\end{figure}

\clearpage
\begin{figure}
% \epsscale{0.6}
\vspace{-1.0in}
\plotone{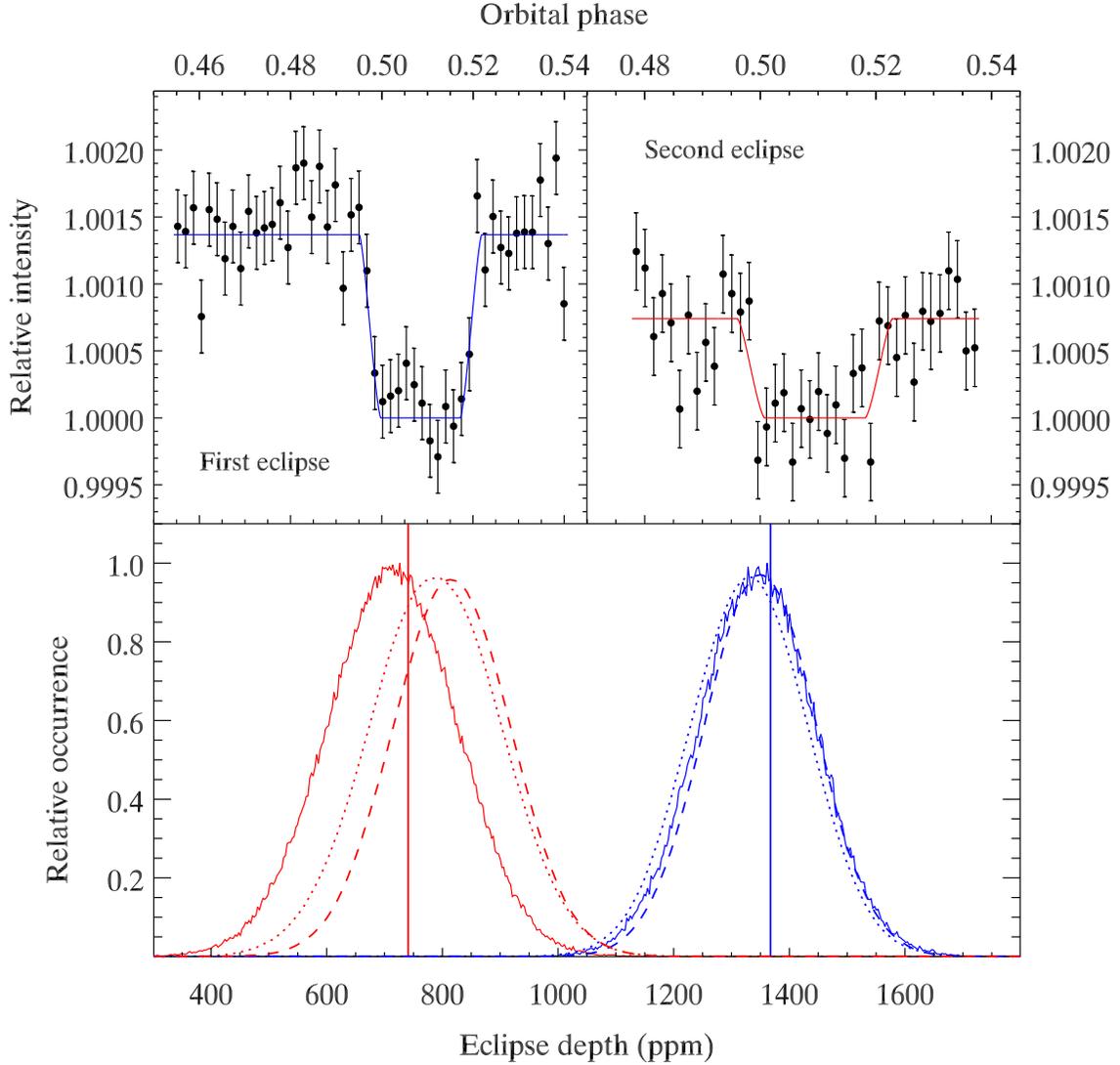}
\vspace{-1.0in}
\caption{Two eclipses of HAT-P-20b at 4.5\,$\mu$m. The observations
  were binned to 50 (first eclipse) and 40 (second eclipse) points per
  data set for clarity of illustration.  Intensity is normalized to
  unity in eclipse (star alone contributing). The lower panel shows
  the posterior distributions for eclipse depth. The distributions
  shown as dashed lines omit the corner pixels from the PLD fit, and
  the distributions shown as dotted lines use a polynomial
  decorrelation, but retaining the broad bandwidth criterion of
  Sec.~3.3.  Both the dashed and dotted distributions have been
  smoothed slightly to make them more legible.  The vertical lines are
  the minimum $\chi^2$ values chosen using our broad bandwidth
  criterion. }
\label{Fig12}
\end{figure}

\clearpage
\begin{figure}
% \epsscale{0.7}
\plotone{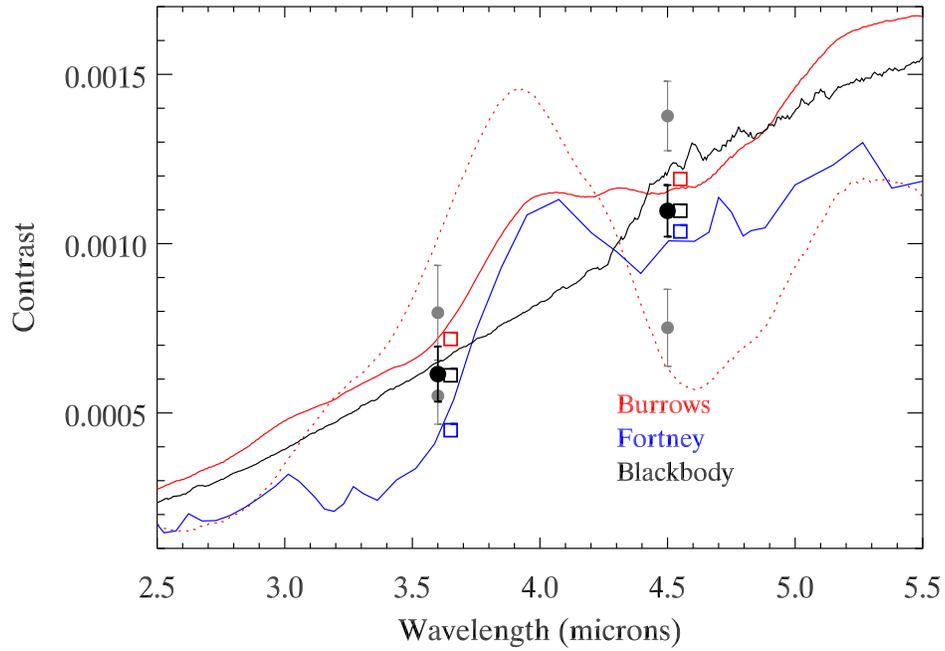}
\vspace{-2.0in}
\caption{Results for HAT-P-20b eclipse depths in the two Warm Spitzer
  bands, averaging both eclipses in each band. The values from
  individual eclipses are plotted in light gray.  The observations are
  compared to the contrast expected for a solar abundance planet
  having day-side re-radiation, with two different models from Adam
  Burrows and Jonathan Fortney.  The black line is a 1134K blackbody.
  We used a Phoenix metal-rich model atmosphere for the star
  (4600/4.5/0.3) \citep{allard}.  The open points show the values
  expected when the stellar and planetary fluxes are integrated over
  the Spitzer bandpass functions; they are offset slightly to longer
  wavelength for visual clarity.  The red dotted line is a Burrows
  model with 10 times solar metallicity for the planet.}
\label{Fig13}
\end{figure}

\clearpage
\begin{figure}
\epsscale{0.6}
%  \vspace{-1.0in}
\plotone{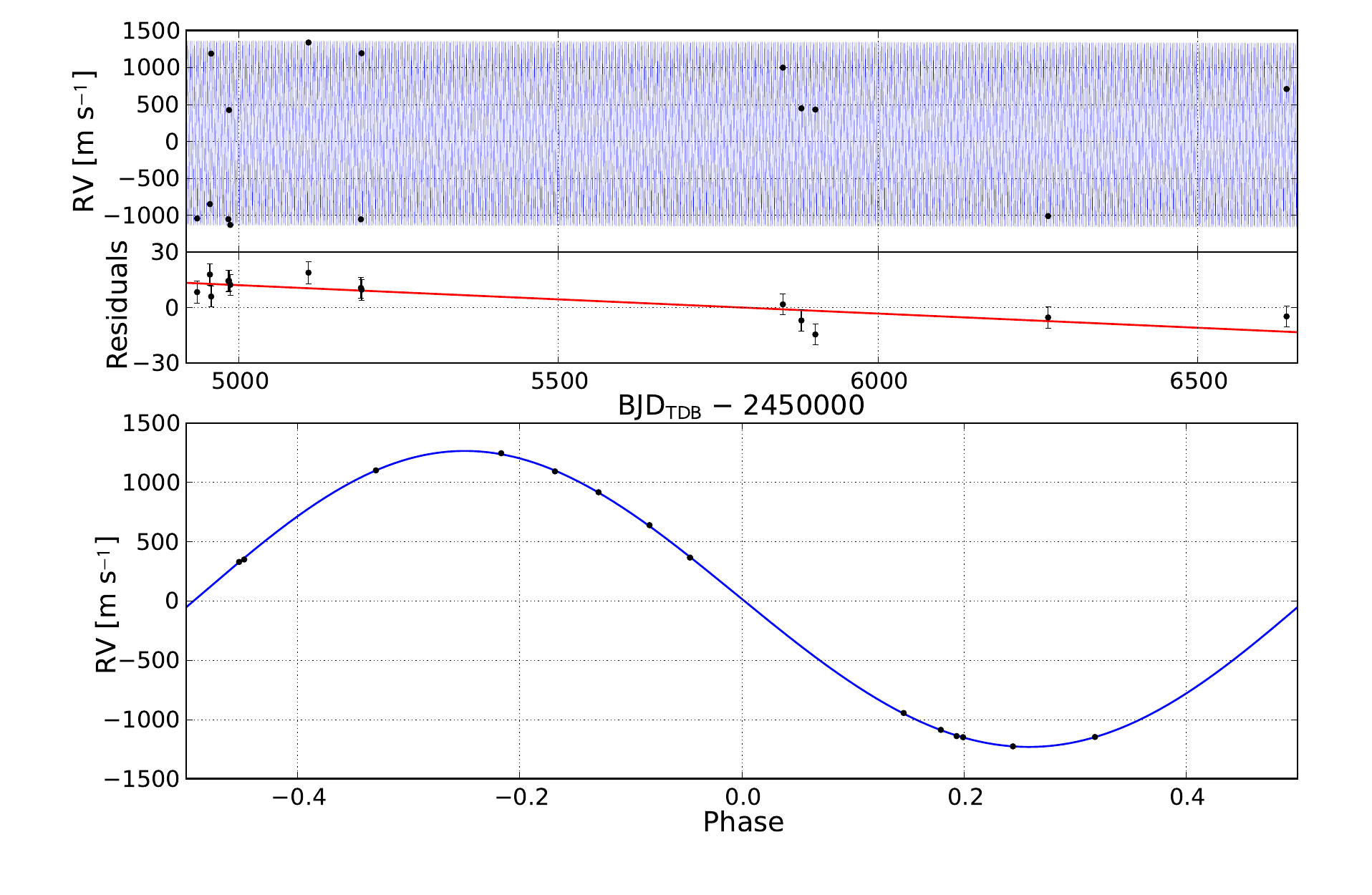}
%  \vspace{1.0in}
\caption{Joint fit of RV data from \citet{knutson14}, and our
  secondary eclipse times for HAT-P-20b. {\it Top:} RV time series for
  HAT-P-20b and best-fit model in blue.  Data are from
  \citet{knutson14}, plus one additional measurement.  {\it
    Top-lower:} Residual from the best-fit single-planet model.  The
  best fitting linear trend has not been subtracted.  The
  statistically significant linear trend first noted in
  \citet{knutson14} is clearly visible, and has continued through the
  most recent data point.  This trend is likely caused by the distant
  stellar companion. {\it Lower:} Radial velocity curve phase-folded
  to the best-fit ephemeris.  Phase 0.0 is the primary transit.}
\label{Fig14}
\end{figure}

%%%%%%%%%TABLES%%%%%%%%%

\begin{sidewaystable}
\begin{center}
  \caption{Parameters used in PLD Fitting, and results for secondary
    eclipses.  Wavelength ($\lambda$) is in microns.  Cen refers to
    the type of centroiding used in the photometry, either 2-D
    Gaussian fitting (G) or center-of-light (C).  Aper is the aperture
    size in IRAC pixels, with V indicating a variable radius
    noise-pixel aperture, with the indicated constant amount added.
    Ramp refers to the time baseline, either quadratic or
    linear. Nbin is the bin size in individual frames that was used in
    the solution.  The cutoff is the earliest orbital phase used in
    each analysis.  SDNR = standard deviation of the normalized
    residuals in the unbinned data. Slope is the slope of the
    $\log{\sigma}$ vs. $\log{N}$ relation used to judge the red noise.
    The eclipse depths and errors are given in parts-per-million
    (ppm).  \label{tbl-1}}
\scalebox{0.8}{
\begin{tabular}{lllllllllll}
\tableline
\tableline
 Eclipse  & $\lambda$ & Cen & Aper & Ramp & Nbin & Cutoff & SDNR & Slope & Eclipse Depth \& Phase & Previous Depth \& Phase \\
\tableline
%      WASP-14  & 3.6  & G  & V+0.4  & Quad & 192  & 0.4107 & 3522   & $-0.510$  & $1925\pm82$  \&  $0.4835\pm 0.0003$ &   \\
 WASP-14  & 3.6  & G  & V+0.0  & Lin  &  10  & 0.4107 & 3054   & $-0.494$  & $1981\pm66$  \&  $0.4833\pm 0.0004$ &  $1900\pm82$ \& $0.4825\pm0.0003$  \citet{blecic}  \\
 WASP-8   & 3.6  & G  & 1.8    & Quad & 148  & 0.49416 & 5414  & $-0.492$  & $906\pm74$  \&  $0.51430\pm0.00019$ & $1130\pm180$ \& $0.51428\pm0.00034$ \citet{cubillos} \\
 WASP-8   & 3.6  & -  &  -     & Quad & 148  & 0.49416 & 5464  & $-0.501$  & $852\pm77$  \&  $0.51453\pm0.00021$ &  sum-of-pixels solution  \\
 CoRoT-2  & 3.6  & G  & V+0.2  & Lin  &  52  & 0.4100 & 6935   & $-0.495$  & $3724\pm200$  \&  $0.4996\pm 0.0008$ & $3550\pm200$ \& $0.4994\pm0.0007$ \citet{deming11} \\
 WASP-12  & 3.6  & C  & V+0.1  & Lin  & 140  & 0.4    & 6864   & $-0.470$  & $4051\pm202$  \&  $0.4985\pm 0.0010$ & $3790\pm130$ \& $0.5010\pm0.0006$ \citet{campo} \\
 WASP-52  & 4.5  & G  & V+0.0  & None & 256  & None   & 7426   & $-0.441$  & Transit $26723\pm200$ \&  0.0    & $27000$ \& 0.0           \\
          &  -   & -  &  -     &  -   &  -   &  -     &  -     &    -      & Eclipse $2365\pm140$ \& 0.5      & $2339$ \& 0.5             \\
          &  -   & -  &  -     &  -   &  -   &  -     &  -     &    -      & Phase curve $951\pm73$          &  $1170$                     \\
 HAT-P-20 & 3.6  & G  & V+0.0  & Lin  & 32   & 0.454  & 2904   & $-0.475$  & $550\pm84$  \&  $0.5090\pm 0.0011$ &   \\
 HAT-P-20 & 3.6  & C  & V+0.2  & Quad & 48   & 0.484  & 3150   & $-0.499$  & $796\pm140$  \&  $0.5078\pm 0.0008$ &    \\
 HAT-P-20 & 4.5  & G  & 2.5    & Quad & 2    & None   & 3963   & $-0.506$  & $1377\pm103$  \&  $0.5084\pm 0.0004$ &    \\
 HAT-P-20 & 4.5  & G  & 3.5    & Lin  & 1    & 0.4778 & 3930   & $-0.485$  & $752\pm114$   \&  $0.5089\pm 0.0008$ &    \\
\tableline
\end{tabular}}
\end{center}
\end{sidewaystable}

\begin{sidewaystable}
\begin{center}
  \caption{Observational parameters for the five real Spitzer data sets used to test and validate our PLD method.  
   $\sigma_{ph}$ is the theoretical noise level for a single exposure, in parts-per-million. $\delta_{pix}$ is the
   total peak-to-peak value of image motion during each data set, in IRAC pixels and corrected for measurement errors. \label{tbl-2}}
\scalebox{0.8}{
\begin{tabular}{lllllllll}
  \tableline
  \tableline
  System & K-magnitude  &  Starting time (UT)  &  AOR number &  Duration (hours) &  Exposure time (sec) &  $\sigma_{ph}$  
   &  $\delta_{pix}$ & Reference/comment \\
  \tableline
  GJ\,436 &  6.07   &  Jan~28,~2010~06:41  &  38702592  &  17.8  &  0.1  & 4880 & 0.23 & \citet{ballard}; no eclipse \\
  CoRoT-2 &  10.31  &  Nov~24,~2009~18:22  &  31774976  &   7.7  &  2.0  & 5650 & 0.09 &  \citet{deming11}    \\
  WASP-14 &  8.62  &  Mar~18,~2010~23:17  &  31760384  &   7.7   &  2.0  & 2600 & 0.14 &  \citet{blecic}  \\
  WASP-8  &  8.09  &  Jul~23,~2010~19:39 &  39200512  &  7.6  &  0.4 & 4700 & 0.07  & \citet{cubillos}  \\
  WASP-12 & 10.19  &  Nov~17,~2010~06:50  & 41260032  &  5.2  &  2.0 & 5440 & 0.13 &  \citet{cowan12} \\
  \tableline
\end{tabular}}
\end{center}
\end{sidewaystable}

\begin{table}
\begin{center}
  \caption{Times of Spitzer observations of HAT-P-20, using IRAC
    subarray mode.  $\sigma_{ph}$ is the theoretical noise level for a
    single exposure, in parts-per-million. $\delta_{pix}$ is the total
    peak-to-peak value of image motion during each data set, in IRAC
    pixels and corrected for measurement errors.  The eclipses are
    listed in the same order as for the HAT-P-20 eclipses in
    Table~1. \label{tbl-3}}
\begin{tabular}{llllll}
  \tableline
  \tableline
  Wavelength ($\mu$m) &  HJD(start)  &  HJD(end) & Number of exposures &  $\sigma_{ph}$  &  $\delta_{pix}$    \\
  \tableline
  3.6       &  2456062.705  &     2456062.954  &  10624 & 2500 & 0.06  \\
  3.6       &  2456810.376  &     2456810.553  &   7552 & 2570 & 0.09  \\
  4.5       &  2456085.719  &     2456085.967  &  10624 & 3500 & 0.06  \\
  4.5       &  2456816.113  &     2456816.289  &   7488 & 3530 & 0.19  \\
  \tableline
\end{tabular}
\end{center}
\end{table}

\begin{table}
\begin{center}
  \caption{Results for HAT-P-20b, for individual eclipses, as well as
    averaging both eclipses at each wavelength, and a grand average
    orbital phase for all four eclipses. The eclipse times are
    BJD(TDB). The phase error for the grand average includes a
    $4.0\times 10^{-6}$ day uncertainty in the orbital period
    \citep{bakos}. The error for the average eclipse depth at
    4.5\,$\mu$m does not include the possible variability in the
    eclipse depth (see text).  \label{tbl-4}}
\begin{tabular}{llll}
\tableline
\tableline
  Wavelength ($\mu$m) &  Eclipse depth (ppm)  &  Eclipse time  &  Eclipse phase \\
\tableline
  3.6     &  $550\pm84$   & $2456062.87458\pm0.00308$  &   $0.5090\pm0.0011$  \\
  3.6     &  $796\pm140$  & $2456810.45414\pm0.00241$  &   $0.5078\pm0.0008$  \\
  4.5     &  $1377\pm103$ & $2456085.87540\pm0.00127$  &   $0.5084\pm0.0004$  \\
  4.5     &  $752\pm114$  & $2456816.20794\pm0.00236$  &   $0.5089\pm0.0008$   \\
  3.6 average      &  $615\pm82$   &     -        &  $0.5082\pm0.0007$  \\
  4.5 average      &  $1096\pm77$  &     -        &  $0.5085\pm0.0005$   \\
  grand average    &     -         &     -        &  $0.50843\pm0.00041$   \\
\tableline
\end{tabular}
\end{center}
\end{table}

    \begin{table}
    \begin{center}
    \caption{HAT-P-20 Orbit Results}
\begin{tabular}{lrr}
\tableline
\tableline
    Parameter &  Value & Units \\ 
\tableline
\\[-5pt]
%  \sidehead{RV Step Parameters}log($P_{b}$) & 0.45868592 $\pm 4.6e-07$ & log(days)
%  {RV Step Parameters}log($P_{b}$) & 0.45868592 $\pm 4.6e-07$ & log(days) \\
$T_{\textrm{conj},b}$ & 2455598.48484 $^{+0.00032}_{-0.0003}$ & \bjdtdb\\
$\sqrt{e_{b}}\cos{\omega_{b}}$ & 0.1035 $^{+0.0049}_{-0.0051}$ & \\
$\sqrt{e_{b}}\sin{\omega_{b}}$ & -0.08 $^{+0.017}_{-0.014}$ & \\
log($K_{b}$) & 3.0959 $^{+0.0011}_{-0.001}$ & \ms\\
$\gamma$ & 83.1 $^{+2.4}_{-2.3}$ & \ms\\
$\dot{\gamma}$ & -0.0154 $^{+0.0037}_{-0.0039}$ & \ms day$^{-1}$\\
$\ddot{\gamma}$ & $\equiv$ 0.0 $\pm 0.0$ & \ms day$^{-2}$\\
jitter & 7.0 $^{+2.2}_{-1.6}$ & \ms\\
% 
%  *********************
% $T_{c,b}$ & 2455080.92738 $^{+0.00021}_{-0.0002}$ & \bjdtdb\\
% $\sqrt{e_{b}}\cos{\omega_{b}}$ & 0.1046 $^{+0.0057}_{-0.006}$ & \\
% $\sqrt{e_{b}}\sin{\omega_{b}}$ & -0.079 $^{+0.018}_{-0.015}$ & \\
% log($K_{b}$) & 3.0958 $^{+0.001}_{-0.0011}$ & \ms\\
% $\gamma$ & 83.4 $^{+2.6}_{-2.5}$ & \ms\\
% $\dot{\gamma}$ & -0.015 $^{+0.0042}_{-0.0041}$ & \ms day$^{-1}$\\
% $\ddot{\gamma}$ & $\equiv$ 0.0 $\pm 0.0$ & \ms day$^{-2}$\\
% jitter & 7.0 $^{+2.3}_{-1.7}$ & \ms\\
% ***********************************
       &                      &    \\
  {RV Model Parameters} &    &    \\
\tableline
\\[-5pt]
$P_{b}$ & 2.8753187 $\pm 1.8e-06$ & days\\
$T_{\textrm{conj},b}$ & 2455598.48484 $^{+0.00032}_{-0.0003}$ & \bjdtdb\\
$e_{b}$ & 0.0171 $^{+0.0018}_{-0.0016}$ & \\
$\omega_{b}$ & 322.4 $^{+7.4}_{-5.9}$ & degrees\\
$K_{b}$ & 1247.0 $^{+3.0}_{-2.9}$ & \ms\\
$\gamma$ & 83.1 $^{+2.4}_{-2.3}$ & \ms\\
$\dot{\gamma}$ & -0.0154 $^{+0.0037}_{-0.0039}$ & \ms day$^{-1}$\\
$\ddot{\gamma}$ & $\equiv$ 0.0 $\pm 0.0$ & \ms day$^{-2}$\\
jitter & 7.0 $^{+2.2}_{-1.6}$ & \ms\\
%
% $P_{b}$ & 2.875318 $\pm 3e-06$ & days\\
% $T_{c,b}$ & 2455080.92738 $^{+0.00021}_{-0.0002}$ & \bjdtdb\\
% $e_{b}$ & 0.0172 $^{+0.0018}_{-0.0017}$ & \\
% $\omega_{b}$ & 322.9 $^{+7.8}_{-6.3}$ & degrees\\
% $K_{b}$ & 1246.9 $^{+2.9}_{-3.0}$ & \ms\\
% $\gamma$ & 83.4 $^{+2.6}_{-2.5}$ & \ms\\
% $\dot{\gamma}$ & -0.015 $^{+0.0042}_{-0.0041}$ & \ms day$^{-1}$\\
% $\ddot{\gamma}$ & $\equiv$ 0.0 $\pm 0.0$ & \ms day$^{-2}$\\
% jitter & 7.0 $^{+2.3}_{-1.7}$ & \ms\\
                         &  &    \\
{RV Derived Parameters}  &   &   \\
\tableline
\\[-5pt]
$e\cos{\omega}$ & 0.01352 $^{+0.00054}_{-0.00057}$ & \\
$e\sin{\omega}$ & -0.0104 $^{+0.0026}_{-0.0025}$ & \\
% $e\cos{\omega}$ & 0.01369 $^{+0.0008}_{-0.00079}$ & \\
% $e\sin{\omega}$ & -0.0104 $^{+0.0027}_{-0.0025}$ & \\
\end{tabular}
\end{center}
%    \enddata
    \tablenotetext{}{Reference epoch for $\gamma$,$\dot{\gamma}$,$\ddot{\gamma}$: 2455787.0}
    \end{table}
%    
%  \begin{table}
%  \begin{center}
%  \caption{Priors for HAT-P-20 Orbit Analysis}
%  \begin{tabular}{lrrr}
%  \tableline
%  \tableline
%  {Parameter} & {Prior} & {Width} & {Units} \\
% \tableline
%    \startdata 
% $P_{b}$ & 2.875317 & 4e-06 & days\\
% $T_{c,b}$ & 2455080.92738 & 0.00021 & \bjdtdb\\
% Ts$_{1}$ & 2456062.8746 & 0.0031 & \bjdtdb \\
% Ts$_{2}$ & 2456810.4541 & 0.0024 & \bjdtdb \\
% Ts$_{3}$ & 2456085.8754 & 0.0013 & \bjdtdb \\
% Ts$_{4}$ & 2456816.2079 & 0.0024 & \bjdtdb \\
% \end{tabular}
% \end{center}
% \end{table}
%
\end{document}